\documentclass[reprint,superscriptaddress,pre,aps,showpacs]{revtex4-1}
\usepackage{dcolumn}
\usepackage{bm}
\usepackage{amsmath}
\usepackage{amsfonts}
\usepackage{amssymb}
\usepackage{bm}
\usepackage[pdftex]{graphicx} 
\usepackage{fullpage}
\usepackage{subfigure}
\usepackage{hyperref} 

\makeatletter

\newcommand{\vast}{\bBigg@{3.5}}

\newcommand{\Vast}{\bBigg@{4}}

\makeatother

\begin{document}

\title{A multi-species model with interconversion, chipping and injection\\}

\author{Himani Sachdeva}
\affiliation{Department of Theoretical Physics, Tata Institute of Fundamental Research,\\ Homi Bhabha Road, Mumbai-400005, India}

\author{Mustansir Barma}
\affiliation{Department of Theoretical Physics, Tata Institute of Fundamental Research,\\ Homi Bhabha Road, Mumbai-400005, India}

\author{Madan Rao}
\affiliation{Raman Research Institute, C.V. Raman Avenue, Bangalore 560080, India}
\affiliation{National Centre for Biological Sciences (TIFR), Bellary Road, Bangalore 560065, India}

\begin{abstract}
Motivated by the phenomenology of transport through the Golgi apparatus of cells, we study a multi-species model with boundary injection of one species of particle, interconversion between the different species of particle, and driven diffusive movement of particles through the system by chipping of a single particle from a site. The model is analysed in one dimension using equations for particle currents. It is found that, depending on the rates of various processes and the asymmetry in the hopping, the system may exist either in a steady phase, in which the average mass at each site attains a time-independent value, or in a ``growing'' phase, in which the total mass grows indefinitely in time, even in a finite system. The growing phases have interesting spatial structure. In particular, we find phases in which some spatial regions of the system have a constant average mass, while other regions show unbounded growth. 
\end{abstract}

\maketitle

\section{Introduction}
\label{sec:intro}
\paragraph*{}
Living cells possess multiple trafficking pathways, which have in common, a regulated flux of cargo molecules, such as proteins and lipids, moving through and processed within, organized compartments or organelles. For instance, the {\it secretory pathway} consists of molecules that move from the endoplasmic reticulum (ER) to the cell surface via an organelle
system called the Golgi apparatus \cite{alberts_molbio}. The Golgi apparatus itself consists of distinct sub-compartments  known as cisternae. Proteins and lipids arrive from the ER, enter through one face of the Golgi and undergo several chemical reactions (processing); the modified products then leave through the other face to mainly target the cell surface. 
\paragraph*{}
There has been much discussion about what drives the flux of molecules through the Golgi \cite{golgi_rev}. The \emph{vesicular transport model} envisages that the cisternae are stable structures with fixed enzymatic composition. Molecules shuttle from one cisterna to the next in small sacs called vesicles, and get chemically modified by the resident enzymes. The \emph{cisternal maturation model}, on the other hand, considers the cisternae to be transient structures that are formed by fusion of incoming vesicles. In this model, it is the cisternae that progress through the Golgi apparatus, carrying the biomolecules with them. Specific enzymes get attached to a cisterna in different stages of its progression, and modify its contents. The final cisterna eventually breaks up, releasing processed biomolecules. Independent studies on a variety of cells provide evidence for both these possibilities \cite{cisternalmat1,cisternalmat2,golgi_exp2}.
\paragraph*{}
Not only the cargo molecules, but indeed the molecules that form the Golgi organelle 
themselves, must also be trafficked along the same route and by the same driving forces. 
This invites the following question: How does the Golgi organelle form in the first place, i.e., how does 
one obtain stable structures (cisternae) given the rules of molecular trafficking, which 
 broadly, may be described as: (i) localised injection of `particles', i.e. of the vesicles containing unprocessed biomolecules (ii) transformation of particles from one species to the other, i.e. chemical processing of the biomolecules by enzymes (iii) transport of particles either by chipping (breaking off) of a single particle (corresponding to vesicle movement) or through movement of bigger aggregates (corresponding to cisternal progression). The aim of this paper is to construct a statistical model incorporating these elementary processes, and use this to quantitatively address questions of structure  formation and the nature of the states at long times. 
\paragraph*{}
With this motivation, we define the following multi-species model. Particles of species A are injected into a one-dimensional (1D) lattice at one boundary. Particles of \emph{different} species B, C... (or more generally, of all types A,B,C etc.) leave from the other boundary. This happens by allowing A particles to convert to B particles (and vice versa), B particles to C and so on. There is no restriction on the number of particles of any species a site can hold. The hopping of particles from one site to another can either occur collectively via movement of the whole stack or one at a time by chipping of a single particle. Chipping refers to a \emph{single} particle breaking off from a stack and hopping to a neighbouring site. The hopping probability may be the same to the left and right (diffusive) or different (driven diffusive, due to an existing chemical or electrical gradient). When a particle (or a collection of particles) hops on to a stack, it merges with the particles already resident on that site. Thus, stacks constantly gain and lose particles. 
\paragraph*{}
This is a generalization of a well studied model of aggregation and chipping \cite{aggregation1, aggregation2}. Earlier studies dealt with a closed system, with a single species of particle. The present generalization deals with an open system, with injection of particles at one end, and interconversion from one species of particle to another. Interestingly, we find that new sorts of phases can arise in some limits.
\paragraph*{}
The parameter space is large. Thus, it is useful to begin with the study of the `chipping only' model, where there is no movement of stacks as a whole. In the remainder of this paper, we will study the model in this limit, with chipping and interconversion rates taken to be constants, independent of the number of A or B or C... particles on the site. With this assumption of constant rates, we find that for some rates, the system fails to achieve steady state in the sense that unbounded growth of mass occurs. Interestingly, even in these growing states, the particle currents at each site are stationary (time independent) after sufficiently long times. The indefinitely growing average mass at a given site arises simply because the particle currents, though stationary, are not balanced at that site \cite{footnote1}. Thus, we call such a state quasi-stationary.
\paragraph*{} 
Although we have defined the model for an arbitrary number of species, from now on we will focus primarily on the two-species model. The multi-species model is a simple generalization of the two-species case and shows qualitatively similar behaviour, as discussed in Sec. \ref{sec:3species}.
\paragraph*{}
The rest of the paper is organised as follows. Section \ref{sec:modelandresults} defines the model precisely, highlights some connections with other models, and briefly discusses the main results of the paper. In Sec. \ref{sec:first_site}, we analyse the behaviour of the first site in detail. In Sec. \ref{sec:asym_lattice}, we study the case of fully asymmetric hopping for a 1D lattice and solve the current equations at each site. In Sec. \ref{sec:continuum}, a continuum limit is taken and results obtained for general asymmetry. In Sec. \ref{sec:3species}, we briefly discuss the generalization of the model to three species. Section \ref{sec:conclusion} contains a discussion of some issues and extensions.

\section{The Model and Results}
\label{sec:modelandresults}
\subsection{Model}
\label{sec:model}
\begin{figure*} 
\includegraphics[width=0.8\textwidth]{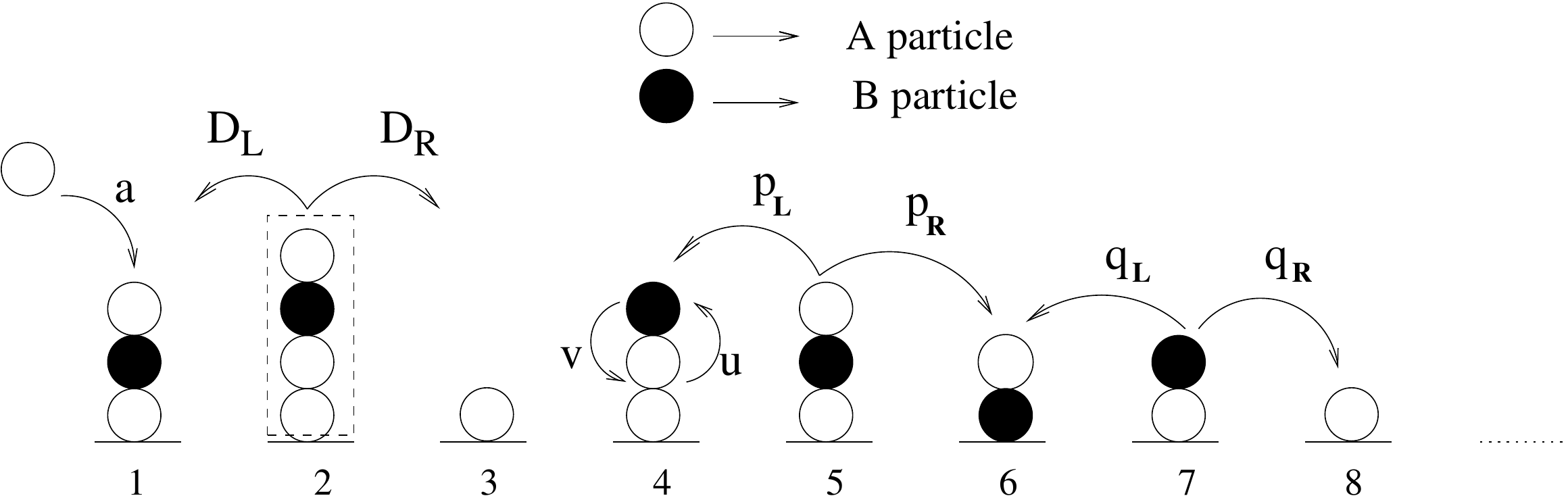}
\caption{Illustration of the allowed moves with A (white) and B (black) particles: Injection of A particles at rate $a$ at site 1. $A\rightarrow B$ and $B\rightarrow A$ conversions at rates  $u$ and $v$ respectively (see site 4). An A particle can hop rightwards (e.g. from site 5 to 6) at rate $p_{R}$ and leftwards (to site 4) at  rate $p_{L}$. Similarly, a B particle (from site 7), can hop to sites 8 or 6 at rates $q_{R}$ and $q_{L}$ respectively. Stack movement corresponds to all particles on a given site (here site 2) collectively hopping to the left or right neighbour (site 1 or 3) at rate $D_{L}$ or $D_{R}$ respectively.}
\label{model_schematic}
\end{figure*}

 The model is defined on a 1D lattice. At any time $t$, a lattice site $i$ has $m^{A}_{i}$ particles of type $A$ and $m^{B}_{i}$ particles of type $B$. We start with an empty lattice with $L$ sites at $t=0$. At each instant, a site $i$ is chosen at random and one of the following moves (illustrated in Fig. \ref{model_schematic}) occurs in an infinitesimal time interval $\Delta t$:
\renewcommand{\theenumi}{\roman{enumi}}
\begin{enumerate}
\item Injection: If the site picked is site 1, then an A particle is injected into it from the left boundary with probability $a\Delta t$: $m_{1}^{A}\rightarrow m_{1}^{A}+1.$
\item Interconversion: With probability $u\Delta t$, one of the A particles residing on the selected site converts to type B: if $m_{i}^{A}>0$, then $m_{i}^{A}\rightarrow m_{i}^{A}-1$ and $m_{i}^{B}\rightarrow m_{i}^{B}+1$. With probability $v\Delta t$ one of the B particles residing on the site converts to type A: if $m_{i}^{B}>0$ then $m_{i}^{B}\rightarrow m_{i}^{B}-1$ and $m_{i}^{A}\rightarrow m_{i}^{A}+1$. 
\item Chipping: With probability $p_{R}\Delta t$ ($p_{L}\Delta t$), one of the A particles chips off the site and hops to the right (left) neighbouring site, thereby decreasing the mass of A on the site by 1 and increasing the mass of A on the neighbouring site by 1; if $m_{i}^{A}>0$ then $m_{i}^{A}\rightarrow m_{i}^{A}-1$ and $m_{i \pm 1}^{A}\rightarrow m_{i \pm 1}^{A}+1$. 
With probability $q_{R}\Delta t$ ($q_{L}\Delta t$), one of the B particles chips off the site and hops to the right (left): if $m_{i}^{B}>0$ then $m_{i}^{B}\rightarrow m_{i}^{B}-1$, $m_{i \pm 1}^{B}\rightarrow m_{i \pm 1}^{B}+1.$ 
\item Stack Movement: With probability $D_{R}\Delta t$ (or $D_{L}\Delta t$), the entire mass (i.e. all A and B particles together) present at the site hops to the right (left) neighbouring site: $m_{i}^{B}\rightarrow 0$, $m_{i}^{A}\rightarrow 0$, $m_{i \pm 1}^{B}\rightarrow m_{i \pm 1}^{B}+m_{i}^{B}$, $m_{i \pm 1}^{A}\rightarrow m_{i \pm 1}^{A}+m_{i}^{A}$. 
\end{enumerate}
At the last site an A (B) particle can hop out of the system at a rate $p_{R} (q_{R})$, i.e. with the same hopping rate as in the bulk. Once particles hop rightwards from the last site $L$, they cannot return to the system. Thus, site $L+1$ acts as a sink.
\paragraph*{}
In general, the drive i.e. the asymmetry in the rates of the three hopping processes -- hopping of stacks, hopping of A particles after chipping and hopping of B particles after chipping, can be different. However, for simplicity, we will take the asymmetry to be the same for all three processes, i.e. $p_{R}/p_{L}=q_{R}/q_{L}=D_{R}/D_{L}$. Then we can parametrize the rates as follows: $p_{R}=\gamma p$, $p_{L}=(1-\gamma)p$; $q_{R}=\gamma q$, $q_{L}=(1-\gamma)q$; $D_{R}=\gamma D$, $D_{L}=(1-\gamma)D$. Here, $\gamma$ is the measure of the asymmetry in hopping and takes on values between 0 and 1.
\paragraph*{}
The moves (i) to (iv) above define a general model in which particles can move via both stack movement and chipping. However, as indicated in section \ref{sec:intro}, in this paper we will analyse the model in the absence of stack movement i.e. in the limit $D=0$.
\paragraph*{}
Several features of our model, including boundary effects, mass movement and coalescence, and two-species coupling have been studied separately in a number of different contexts.
Macroscopic aggregates are known to form in models such as the the zero range process (ZRP) \cite{ZRP_review} and the aggregation-chipping model \cite{aggregation1,aggregation2}. In these models, the closed system in steady state can enter a condensate phase in which a finite fraction of the total mass resides on a single site. A quasi-stationary state characterised by growth of total mass in the system occurs in aggregation models such as the Takayasu model \cite{Takayasu_model} in which particles are injected at every site. The in-out model \cite{aggregation1,inout}, which extends the Takayasu model to allow for evaporation of unit masses, exhibits a phase transition from a growing state to a steady state. Such a phase transition is also seen in our model. However, the phases in our model are not spatially uniform as injection occurs only at the boundary.
\paragraph*{}
Open boundaries with injection can induce interesting features such as boundary-driven phase transitions in the asymmetric simple exclusion process (ASEP) \cite{ASEP_Schutz}. In the ZRP with open boundaries \cite{ZRP_open}, steady state, in the sense of constant average mass, is not reached for strong boundary injection; indefinitely growing aggregates may form on one or both boundaries, while bulk sites attain steady state or show slower growth, depending on the injection, hopping rates, and asymmetry. Our model, in which injection takes place at the boundary, also shows growing phases; somewhat counter intuitively, a steady state with constant mass is reached close to the boundary, while there is unbounded growth of the mass in the bulk.
\paragraph*{}
Coupling between different species of particle may or may not involve interconversion. For example, in the two-species ZRP \cite{twospZRP1,twospZRP2}, condensates of one or both species may form above critical densities but the number of particles of each species in the system does not change. New effects are seen when interconversion (or more generally, non-conservation of particle number) occurs. Multi-species models that allow for inter-conversion have been studied earlier in the context of transport in single \cite{levine,chou,muhuri,mrao,urna} and multiple \cite{kolomeisky1,kolomeisky2,frey,juhasz} channels\cite{footnote2}. These models differ from the present work in that they restrict the occupancy to at most one particle per site in each channel. Generically in an open boundary channel, if the number of particles is not conserved in the bulk (e.g. if switching between channels or deposition and evaporation occurs), then for low switching rates, the system may develop a localised shock separating low density and high density regions in one of the channels \cite{frey,juhasz}. Our model too shows co-existence of two  phases of one species (a steady phase and a growing phase) in different spatial regions separated by a domain wall with a tunable location.
\subsection{Results}
\label{sec:results}
\begin{figure*} 
\centering
\subfigure[]{
\includegraphics[width=0.45\textwidth]{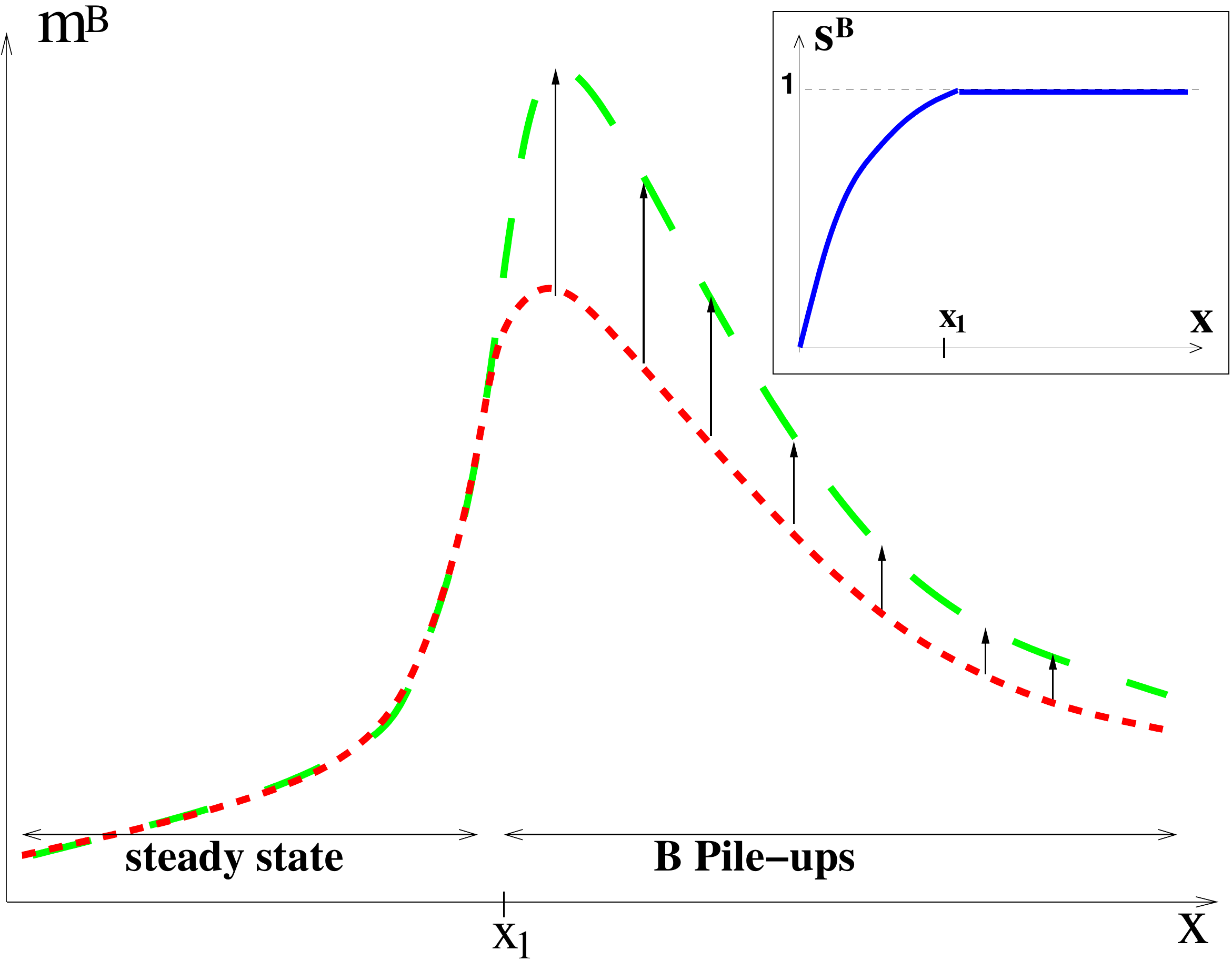}
\label{schematic_asym}}
\subfigure[]{
\includegraphics[width=0.45\textwidth]{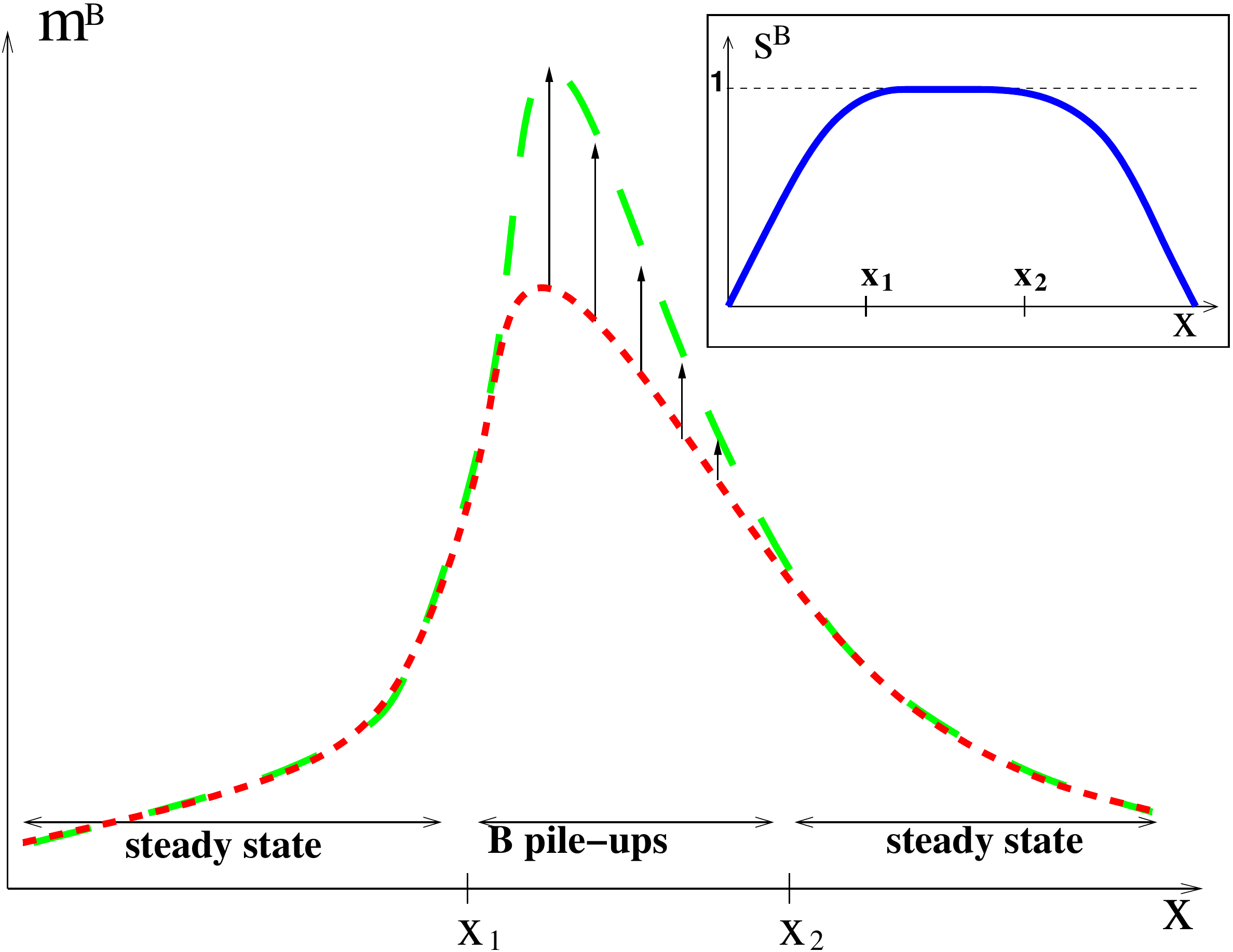}
\label{schematic_sym}}
\caption{Schematic depiction of a phase with pile-ups of B for (a) fully asymmetric ($\gamma=1$) (b) fully symmetric ($\gamma=1/2$) hopping. Dashed lines show $\langle m^{B}\rangle$ as a function of $x$ at two time instants $t_{1}$ (small dashes) and $t_{2}$ (big dashes) respectively such that $0\ll t_{1}<t_{2}$. The two lines overlap in regions of steady state (i.e. $\langle m^{B}\rangle$ remains constant in time); the $t_{2}$ line lies above the $t_{1}$ line in regions of B pile-ups ($\langle m_{B}\rangle$ increases with time). Insets: Occupation probability $s^{B}$ as a function of $x$. While $s^{B} = 1$ in regions of B pile-ups, $s^{B}<1$ in regions of steady state. For $\gamma=1/2$, the region of pile-ups ($x_{1}<x<x_{2}$) exists between regions of steady state near each boundary; for {$\gamma=1$}, the entire region to the right ($x>x_{1}$) of the steady state region has B pile-ups.}
\end{figure*}

Our main results may be summarised as follows:
\\The model is parametrized by the rates $a,p,q,u,v$ and the asymmetry parameter $\gamma$, where $\gamma=1/2$ corresponds to symmetric and $\gamma=1$ to fully asymmetric rightward hopping. In different regions of parameter space, different kinds of behaviour are seen:
\begin{enumerate}
\item For some values of parameters, the system exists in a steady phase. Occupation probabilities of A and B at a site $i$, i.e. the probability that the site has at least one A (B) particle are defined as
\begin{eqnarray}
 s_{i}^{A}(t)= 1-\displaystyle\sum\limits_{m^{B}=0}^\infty P_{i}(0,m^{B},t)\nonumber\\
 s_{i}^{B}(t)= 1-\displaystyle\sum\limits_{m^{A}=0}^\infty  P_{i}(m^{A},0,t)
\label{eqn:sAsB}
\end{eqnarray}
 where $P_{i}(m^{A},m^{B},t)dt$ is the probability of finding $m^{A}$ A particles and $m^{B}$ B particles on site $i$ between time $t$ and $t+dt$. In the steady phase, $s_{i}^{A}$ and $s_{i}^{B}$ at each site, reach a time independent value \emph{less than 1}. The total mass in the system (and on each site) also reaches a finite time-independent value. 
\item For other values of parameters, the system exists in a quasi-stationary state characterised by stationary (time-independent) currents but \emph{non-stationary i.e. indefinitely growing total mass}. This happens even in a finite system. We will refer to unbounded growth of mass as \emph{formation of pile-ups}. More precisely, if the average A (B) mass at a site grows indefinitely, we refer to this as an A (B) pile-up at the site. At the site with an A (B) pile-up, the mean occupancy $s^{A} (s^{B})$ approaches $1$ at long times (see Sec. \ref{sec:first_site}). This may be treated as a functional definition of a pile-up. A pile-up is to be distinguished from a condensate of the sort found in the Zero Range Process (ZRP) \cite{ZRP_review}: A condensate at a site contains a finite fraction of the total number of particles in the system but the condensate mass does not grow in time. 
\item In the growing phase, a system may have pile-ups of A, of B or of both species. For the situation with injection of only A particles, as studied in this paper, an A pile-up can be found only on site 1 (except in the special case $\gamma =1$, $q=0$), i.e. it is a boundary phenomenon. B pile-ups, on the other hand, are found in the bulk. In a phase with B pile-ups, an interesting situation can occur where the lattice has a \emph{region in steady state (with finite mean mass) co-existing with a region of B pile-ups (where the mean mass grows indefinitely)}. For fully asymmetric hopping, the region close to the left boundary is in steady state while pile-ups occur in the entire region to the right (shown schematically in Fig \ref{schematic_asym}). For the symmetric or partially asymmetric case, there are two regions in steady state close to either boundary, with a region of B pile-ups between these (Fig. \ref{schematic_sym}).
\end{enumerate}

In the rest of the paper, we will adopt the following terminology. If the average mass of both A and B reaches a finite time-independent value, we will refer to the system as being in a steady phase. If one region of the system attains constant average mass and another region shows unbounded growth, we will refer to the two regions as a region in steady state and a region with pile-ups respectively. The system as a whole will be referred to as being in a growing phase.

\begin{figure} [h!]
\centering
\subfigure[]{
\includegraphics[width=0.4\textwidth]{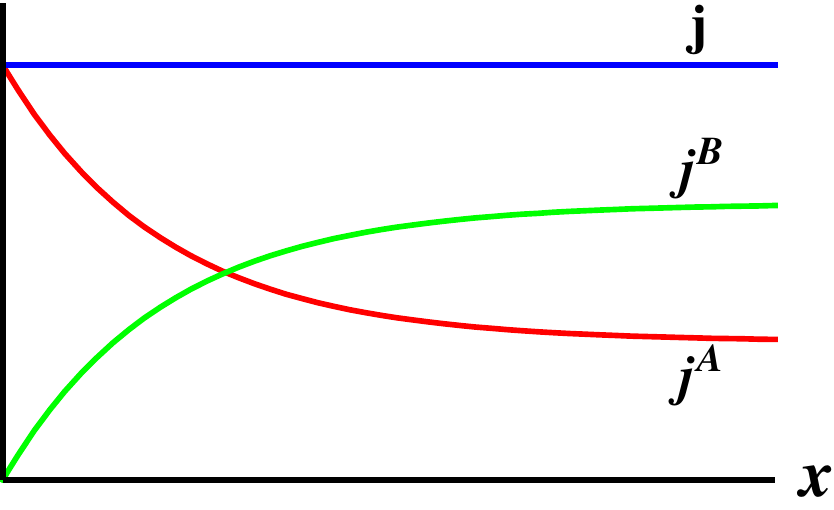}
\label{j(x)steady}}
\subfigure[]{
\includegraphics[width=0.4\textwidth]{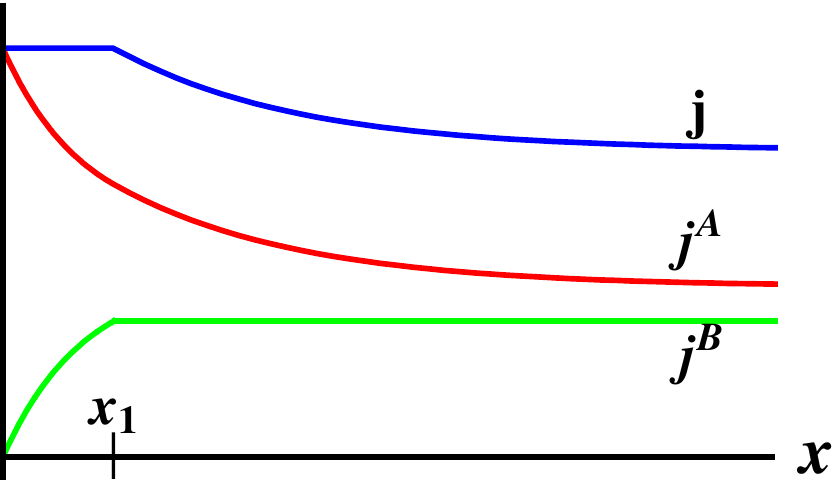}
\label{j(x)growing}}
\caption{Schematic diagram showing variation of the particle currents $j^{A}$, $j^{B}$ and $j$ with distance $x$ from left boundary for $\gamma = 1$ for a) a system in steady phase b) a system with a steady state region and a B pile-up region. In a region of steady state, total particle current $j$ does not change with $x$. In a region of pile-ups, $j(x)$ decreases with $x$ }
\label{jprofile}
\end{figure}
\paragraph*{}
Steady and growing phases can also be distinguished by the spatial profile of the currents. Let $j^{A}_{i}$, $j^{B}_{i}$ and $j_{i} = j^{A}_{i} + j^{B}_{i}$ denote respectively the net A particle current, the B particle current and the total particle current in the $i$th bond of the lattice i.e. from site $i$ to $i+1$. Figs. \ref{j(x)steady} and \ref{j(x)growing} show schematically how, for $\gamma=1$, these three currents vary across the lattice in a steady phase and a growing phase (with B pile-ups) respectively. In the steady phase, although $j^{A}$ and $j^{B}$ vary with  the distance from the origin $x$, they do so in such a way that the total particle current $j$ remains constant across the lattice. In contrast, in the growing phase, the \emph{total particle current $j$ also varies with $x$} in the pile-up region. In fact, $j_{i-1}$ must be greater than $j_{i}$ for a pile-up to exist at site $i$, implying that $j$ decreases with $i$ (or $x$) in a pile-up region.

\begin{figure}
\vspace{-3mm}
\centering
\includegraphics[width=0.4\textwidth]{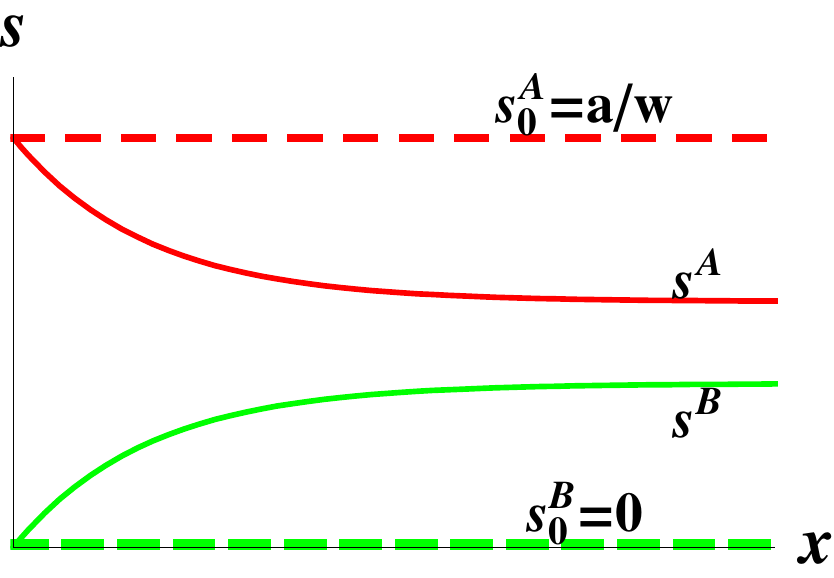}
\caption{In the absence of interconversion ($u=v=0$), the occupation probability profiles $s_{0}^{A}$ and $s_{0}^{B}$ (dashed lines) are flat, being $a/w$ and $0$ respectively. When $u\neq 0$, $v\neq 0$, there is a net interconversion from A to B at each site, pushing $s^{A}$ down from the value $s_{0}^{A}$ and pushing up $s^{B}$ from the value $s_{0}^{B}$. Here $s^{A}$ and $s^{B}$ (solid lines) are occupation probabilities with interconversion.}
\label{s(x)}
\end{figure}

\paragraph*{}
In order to explain the spatial structure of the various phases, it is useful to understand what causes the spatial variation in $j^{A}$ and $j^{B}$ or equivalently, $s^{A}$ and $s^{B}$. For simplicity, let us consider purely rightward hopping ($\gamma=1$) in a system of size $L$. In the absence of interconversion, the particle current $j^{A}$ is the same in each bond, and so is $j^{B}$. Thus, $s^{A}$ and $s^{B}$ are constant across the lattice, being $a/w$ and $0$ respectively (dotted lines in Fig. \ref{s(x)}). Interconversion changes $j^{A}$ and $j^{B}$ and consequently $s^{A}$ and $s^{B}$ (solid lines in Fig. \ref{s(x)}) in the following way.  The particle current into site 1 ($j_{0}$) is entirely of type A. At site 1, some of the A particles convert to type B, resulting in a small probability that site 1 also has B particles. Hence, $j_{1}$ is different in composition, having a small B component as well. At site 2, there is still a much larger concentration of A particles than B particles, and hence a net conversion of A to B particles. Thus, the composition of $j_{2}$ shifts further towards B. 
This process continues until we reach a site at which there is \emph{no net conversion}. Beyond this site $j^{A}$ and $j^{B}$ do not change with $x$. If interconversions are faster than hopping (i.e. $u,v \gtrsim p,q$), then at a given site, the relative amounts of A and B can alter significantly through interconversion before hopping occurs. In this limit, $j^{A}$ and $j^{B}$ change sharply over a few sites and reach their asymptotic values very close to the left boundary. In the other limit ($u,v\ll p,q$), $j^{A}$ and $j^{B}$ vary slowly over a large part of the lattice. In particular, if $u$ and $v$ are not $\ensuremath{\mathcal{O}\!(1)}$ but $\ensuremath{\mathcal{O}\!\left(1/L\right)}$, then the spatial variation in $j^{A}$ and $j^{B}$ becomes a bulk rather than boundary effect. 
\paragraph*{}
The right boundary is an absorbing boundary. The degree of asymmetry in the hopping determines how far into the lattice the effect of the absorbing boundary condition propagates. For fully asymmetric dynamics ($\gamma=1$), the right boundary has no effect on any other site while for $\gamma$ slightly less than 1, the effect extends only into a very small region of the lattice close to the boundary. For $\gamma=1/2$, the information about the right boundary propagates into the entire lattice. Thus, the asymmetry in hopping rates and the rate of interconversion relative to the rate of hopping together determine how $j^{A}$ and $j^{B}$ or $s^{A}$ and $s^{B}$ vary across the lattice.

\section{Analysis of the first site}
\label{sec:first_site}
\paragraph*{}
For fully asymmetric (rightward) hopping ($\gamma=1$), a site is not affected by any sites to the right of it. Thus, the first site can be analysed independently of the rest of the lattice; this analysis yields considerable insight into the full problem. We find four possible behaviours: steady state; pile-up of A but not B; pile-up of B but not A; pile-ups of both A and B.

The master equation for the probability distribution $P(m^{A},m^{B})$ (as defined in section \ref{sec:results}) at the first site can be written as follows:
\\
\begin{subequations}
\label{eqn:mastereqsite1}
\\For $m^{A} \geq\ 1$ and  $m^{B} \geq\ 1$
\begin{equation}
\begin{split}
&\frac{\partial P(m^{A},m^{B})}{\partial t} =pP(m^{A}+1,m^{B})\\
& + qP(m^{A},m^{B}+1) + aP(m^{A}-1,m^{B})\\
& + uP(m^{A}+1,m^{B}-1) + vP(m^{A}-1,m^{B}+1)\\
 &- (a+u+v+p+q)P(m^{A},m^{B})
\end{split}
\end{equation}
\begin{equation}
\begin{split}
& \frac{\partial P(0,m^{B})}{\partial t} = pP(1,m^{B}) + qP(0,m^{B}+1)\\
& + uP(1,m^{B}-1) - (a+v+q)P(0,m^{B})
\end{split}
\end{equation}
\begin{equation}
 \begin{split}
&\frac{\partial P(m^{A},0)}{\partial t} = pP(m^{A}+1,0) + qP(m^{A},1)\\
& + aP(m^{A}-1,0) + vP(m^{A}-1,1) \\
&- (a+u+p)P(m^{A},0)
\end{split}
\end{equation}
\begin{equation}
\frac{\partial P(0,0)}{\partial t} = pP(1,0) + qP(0,1) - aP(0,0)
\end{equation}
\end{subequations}
\\

To understand the behaviour of the system, we do not solve the master equation but work with the time-evolution equations for $\langle m^{A} \rangle$ and $\langle m^{B}\rangle$:
\begin{subequations}
\label{eqn:currenteqsite1}

\begin{equation}
 \frac{d\langle m^{A}\rangle}{dt} = a + vs^{B} - us^{A} - ps^{A}
\end{equation}
\begin{equation}
 \frac{d\langle m^{B}\rangle}{dt} = us^{A} - vs^{B}-qs^{B}
\end{equation}
\end{subequations}
Here $\langle m^{A}\rangle$ and $\langle m^{B}\rangle$ denote the average masses of A and B respectively on site 1 averaged over time histories and $s^{A}$ and $s^{B}$ the occupation probabilities of A and B (defined in Eq. \eqref{eqn:sAsB}); all of these are functions of time. Evidently, Eq. \eqref{eqn:currenteqsite1} is just a continuity equation for the average masses.
\begin{figure*} 
\centering
\subfigure[]{
\includegraphics[width=0.4\textwidth]{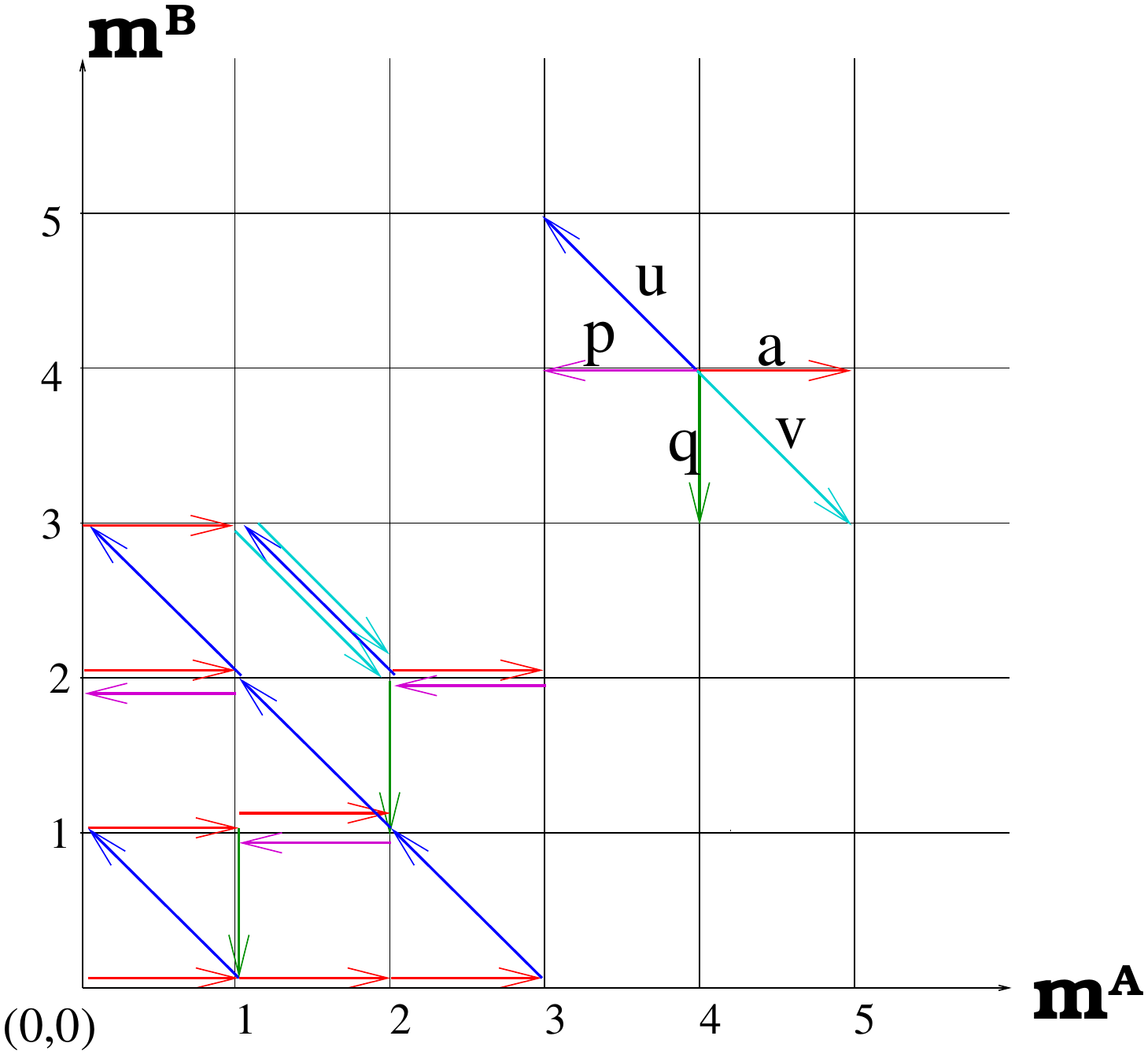}
\label{randomwalk}}
\subfigure[]{
\includegraphics[width=0.4\textwidth]{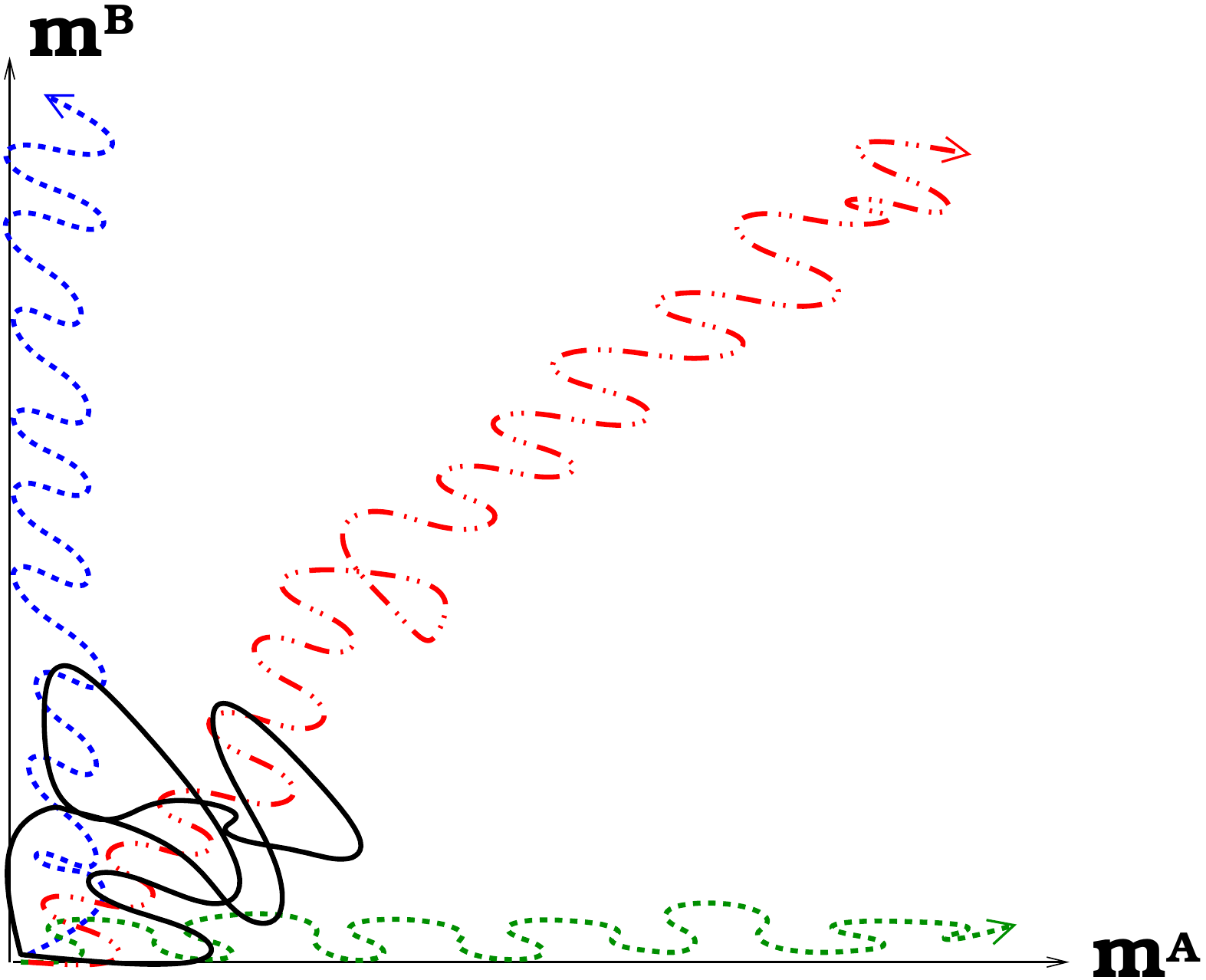}
\label{trails}}
\caption{Time evolution on the first site can be mapped to random walk in 2D $m^{A}-m^{B}$ space. (a) A possible random walk in $m^{A}-m^{B}$ space. Allowed moves are enumerated in the upper right corner of the figure along with the corresponding rates at which they occur.\newline (b) Trails in $m^{A}-m^{B}$ space corresponding to various scenarios. Trail in solid line shows steady state, trail (in dashed line) directed along $m^{A}$ axes shows pile-up of A, trail (in dashed line) along $m^{B}$ axes shows pile-up of B, fourth trail (dot-dash) shows pile-up of both A and B.}

\end{figure*}

\paragraph*{}
The time-evolution of the first site can also be mapped to a random walk in two dimensions. The two-dimensional (2D) space here is the $m^{A}-m^{B}$ space. The random walk starts at the origin at $t=0$, moves in both positive and negative directions along the $m^{A}$ axis (with rates $a$ and $p$ respectively) but moves only in the negative direction along the $m^{B}$ axis (with a rate $q$) as there is no injection of B particles. Diagonal moves, corresponding to interconversion, are also allowed \emph{but only in the direction of increasing $m^{A}$ and decreasing $m^{B}$ or vice versa}. The $m^{A}$ and $m^{B}$ axes are reflecting boundaries. A possible random walk, with all allowed moves in $m^{A}-m^{B}$ space, is illustrated in Fig. \ref{randomwalk}. Clearly, the random walk cannot be decomposed into independent walks along the $m^{A}$ and $m^{B}$ axes because of the diagonal moves.

\begin{figure*}
\subfigure[]{
\includegraphics[width=0.4\textwidth]{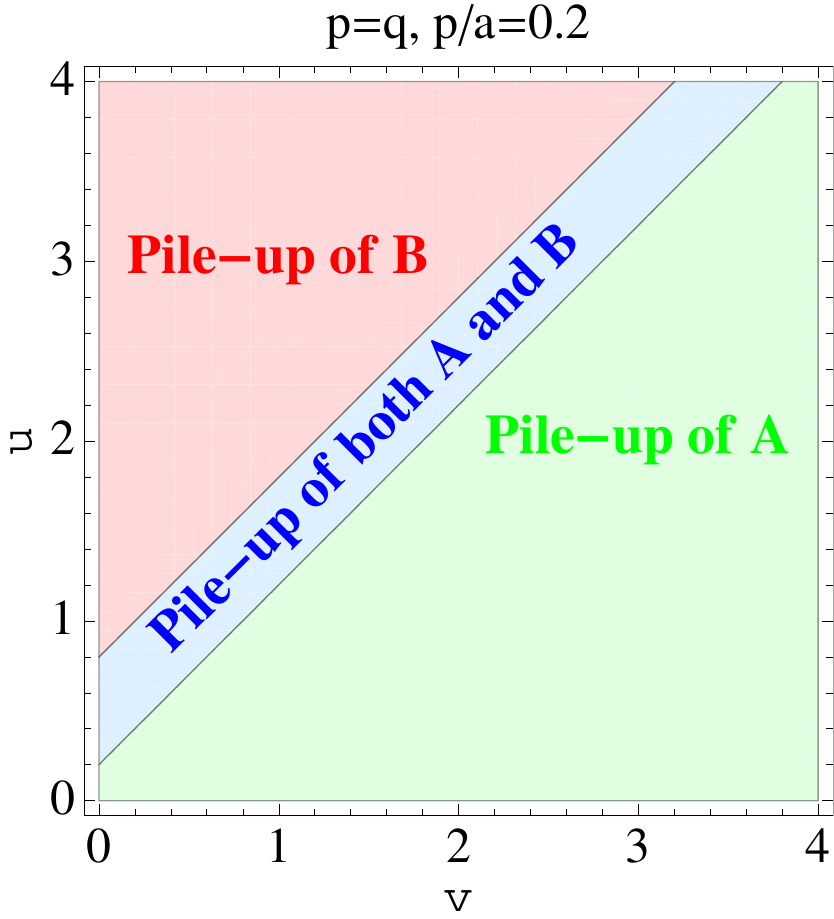}
\label{same_hoppingrates2}}
\subfigure[]{
\includegraphics[width=0.4\textwidth]{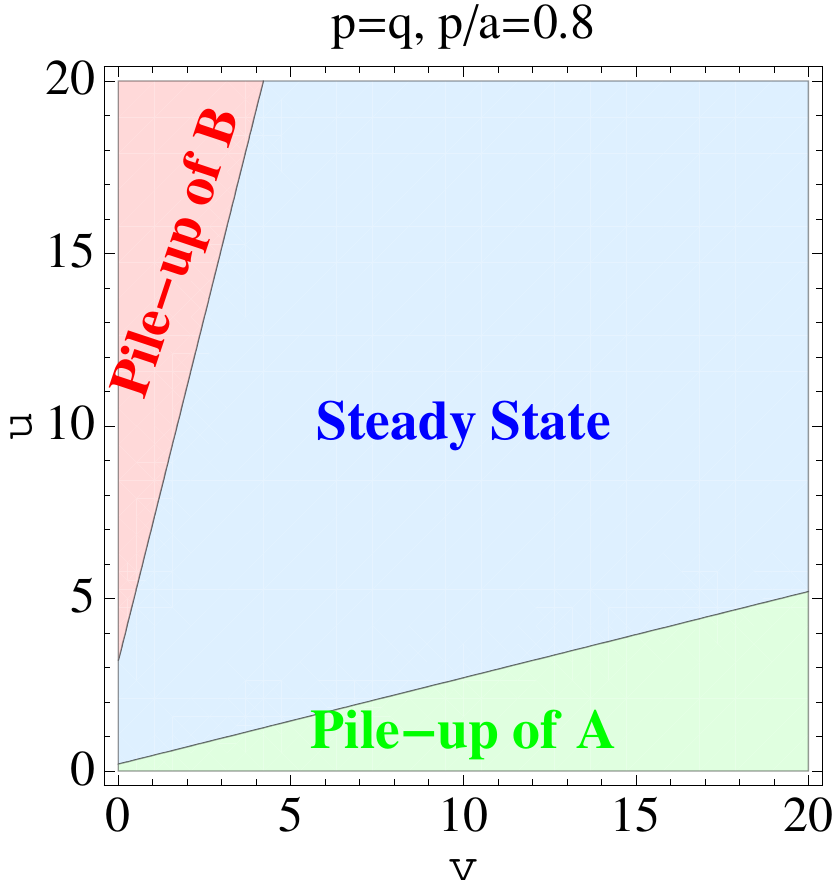}
\label{same_hoppingrates1}}

\caption{$v-u$ phase diagram for the first site when A and B hopping rates are equal
(a) $p=q, p/a=0.2$ (b) $p=q, p/a=0.8$}
\end{figure*}

\paragraph*{}
The random walk picture is useful in the following way. To begin with, let us consider a simple 1D random walk in the positive half of the $x$-axis with a reflecting barrier at the origin. 
This walk has a drift velocity $c$ and a diffusion constant $D$. After a length  $\sim D/|c|$, drift becomes more important than diffusion. If $c>0$, then beyond this length, the mean displacement will grow linearly with time, while if $c<0$, the mean displacement reaches a constant value $\sim D/|c|$. 
If $c=0$, then the motion is always diffusive, and the mean displacement grows as $\sqrt{t}$. The 2-d walk is more complicated because motion along $m^{A}$ and $m^{B}$ axes is coupled due to the diagonal moves (corresponding to interconversion). 
The diagonal moves  effectively make $c$ and $D$ in the $m^{A}$ direction dependent on the average displacement along the $m^{B}$ direction. If the effective $c$ in the $m^{A}$ direction is greater than or equal to zero, then we get a pile-up of A on the first site (with $c=0$ implying $\sqrt t$ growth of the pile-up and $c>0$, growth that is linear in time). Also, as expected from the theory of random walks, the fluctuation about the mean mass, i.e. $\sqrt{\langle(m^{A})^{2}\rangle-\langle m^{A}\rangle^{2}}$ grows as $\sqrt{t}$ (for $c\geq 0$). If $c<0$, then the mean mass of A and the mean deviation in mass of A reach time independent values. In terms of the occupation probability, if $s^{A}<1$, then $m^{A}$ must be finite and if $s^{A}=1$, $m^{A}$ is infinite at long times (i.e. pile-ups occur).
To see this, note that the return from $m^{A}=M$ to $m^{A}=0$ requires $M$ steps, implying that $P(m^{A}=0)$ falls as $e^{-\alpha M}$, which is non-vanishing only for finite $M$. So, if $s^{A}=1-P(m^{A}=0)$ is less than 1 at long times, $<m^{A}>$ must be finite; if $s^{A}$ is equal to 1 at long times, $<m^{A}>$ is infinite.
\paragraph*{}
From the above discussion it is clear that depending on the rates, the effective $c$ in either direction (i.e. along $m^{A}$ and $m^{B}$ axes) can be negative or non-negative, resulting in four possibilities altogether. Figure \ref{trails} shows schematically, possible random walks corresponding to the four scenarios -- steady state, pile-up of A, pile-up of B and pile-up of both A and B.

\paragraph*{}
In terms of the rates, the conditions for each of the four scenarios can be derived from Eq. \eqref{eqn:currenteqsite1}. For example, if there is a pile-up of B but not of A on site 1, then at long times, $d\langle m^{B}\rangle/dt \geq 0$ with $s^{B}=1$ and $d\langle m^{A}\rangle/dt=0$ with $s^{A}<1$. Substituting these into Eq. \eqref{eqn:currenteqsite1} gives the two inequalities in parameter space that must be satisfied in this case. Appendix \ref{appendixA} lists the necessary conditions in parameter space for each of the four cases.

\paragraph*{}
These conditions can be visualised better by considering a reduced parameter space. We look at a case where both species hop with the same rate, i.e. $p=q$. Then the total mass evolves according to:
\begin{equation}
 \frac{d(\langle m^{A}\rangle+\langle m^{B}\rangle)}{dt} = a - p(s^{A}+ s^{B})
\end{equation}
Clearly, if $a<p$, then $s^{A}$, $s^{B} <1$ and the system will always attain steady state. If $a \geq 2p$, the system can never attain steady state and must show pile-ups (as $d(\langle m^{A}\rangle +\langle m^{B}\rangle)/dt\geq0$ even when $s^{A}$ and $s^{B}$ take on the maximum possible value i.e. $1$). In this case, if $u \sim v$, i.e. the interconversion rates are similar, both A and B pile up. However, if one of the conversion rates is much higher than the other, then the species which converts fast is able to stabilize, so that only one of the species piles up. Figure \ref{same_hoppingrates2} shows the $v-u$ phase diagram for the case $a \geq 2p$. If $p \leq a <2p$, then $s^{A}$ and $s^{B}$ can adjust themselves such that steady state is attained. However, as before, if one of the species converts much faster than the other, it drives the latter out of steady state, causing it to pile up. Figure \ref{same_hoppingrates1} shows the $v-u$ phase diagram for the case $p \leq a <2p$.

\section{Asymmetric Hopping on a 1D lattice}
\label{sec:asym_lattice}
\paragraph*{}
In this section, we analyse the fully asymmetric case ($\gamma=1$). In this limit, the current equations involving $s^{A}$ and $s^{B}$ have a simple solution. Also, this case illustrates in a clear way, how interconversions drive  spatial gradients in A and B particle currents. 
\paragraph*{}
Current equations similar to Eq. \eqref{eqn:currenteqsite1} can be written for each site in the system:
\begin{subequations}
\label{eqn:currenteqsitei}
\begin{equation}
 \frac{d<m_{i}^{A}>}{dt} = ps_{i-1}^{A} + vs_{i}^{B} - us_{i}^{A} - ps_{i}^{A}
\end{equation}
\begin{equation}
 \frac{d<m_{i}^{B}>}{dt} = qs_{i-1}^{B} + us_{i}^{A} - vs_{i}^{B}-qs_{i}^{B} \qquad \qquad i\neq 1
\end{equation}
\end{subequations}

and for site $1$, as before: 
\begin{subequations}
\label{eqn:currenteqsite1A}
\begin{equation}
 \frac{d<m_{1}^{A}>}{dt} = a + vs_{1}^{B} - us_{1}^{A} - ps_{1}^{A}
\end{equation}
\begin{equation}
 \frac{d<m_{1}^{B}>}{dt} = us_{1}^{A} - vs_{1}^{B}-qs_{1}^{B} 
\end{equation}
\end{subequations}

The above equations are exact, being just the continuity equations for the average masses. They can also be obtained by writing a master equation for the total probability $\mathcal{P}(\{m_{i}^{A}\},\{m_{i}^{B}\})$, multiplying it by $m^{A}_{i}$ (or $m^{B}_{i}$) and averaging over all configurations.
\subsection{Steady Phase:}

For the steady phase ($d\langle m^{A}\rangle/dt = d\langle m^{B}\rangle/dt=0$ at all sites), equations \eqref{eqn:currenteqsitei} and \eqref{eqn:currenteqsite1A} give recursion relations expressing $s_{i}^{A}$ and $s_{i}^{B}$ in terms of $s_{i-1}^{A}$ and $s_{i-1}^{B}$. Iterating these recursion relations gives the following steady phase occupation probabilities:

\begin{equation}
\label{eqn:ssprofile}
\begin{split}
 &s_{i}^{A}=\frac{av}{uq+vp}\left[1+\left(\frac{uq}{vp}\right)\lambda^i\right] \\
& s_{i}^{B}=\frac{au}{uq+vp}\left[1-\lambda^i\right]
\end{split}
\end {equation}
where $\lambda=(pq)/(uq+pv+pq)$.

\begin{figure}[h]
\centering
\vspace{-8mm}
\includegraphics[width=0.45\textwidth]{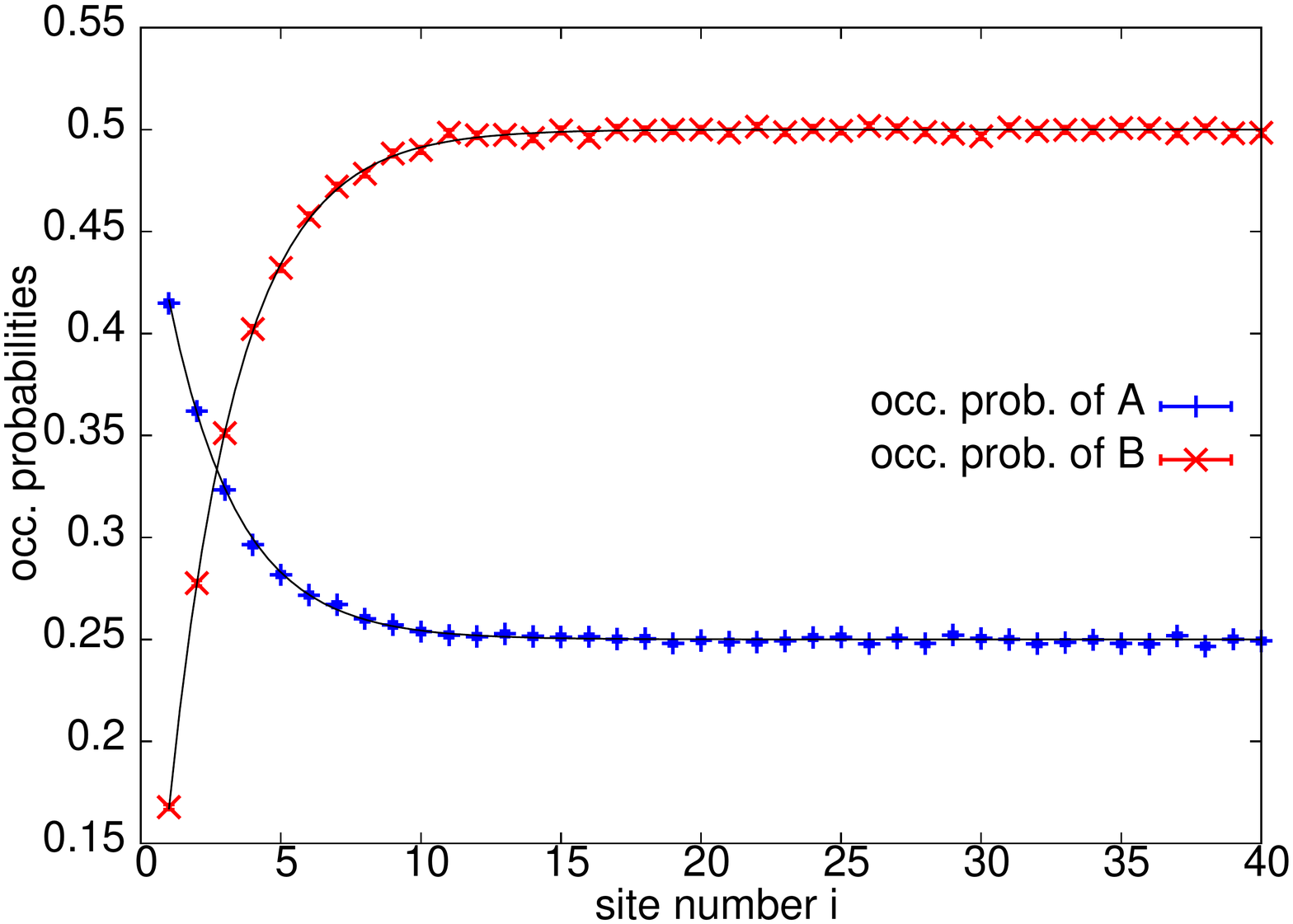}
\caption{Occupation probabilities $s_{i}^{A}$ and $s_{i}^{B}$ as a function of site number $i$ in a steady phase for $\gamma=1$, $a=1$, $u=0.5$, $v=0.25$, $p=2$, $q=1$. Data points are obtained from Monte Carlo simulations, and solid lines are plots of the analytical results in Eq. \eqref{eqn:ssprofile}. $s^{A}$ and $s^{B}$ approach their asymptotic values over a length $l_{D}$ defined in Eq. \eqref{eqn:decay_length}.}
\label{steadystate}.
\label{sstate}
\end{figure}

Figure \ref{sstate} shows the spatial profile of $s_{i}^{A}$ and $s_{i}^{B}$ in a steady phase, as obtained from Monte Carlo simulations, along with the analytical results, for a certain choice of parameters. 
\paragraph*{}
From Eq. \eqref{eqn:ssprofile}, the following can be deduced:
\begin{enumerate}
\item In a steady phase, occupation probabilities $s^{A}_{i}$, $s_{i}^{B}$ must be less than $1$ for all $i$. This is ensured if the maxima of the two expressions in \eqref{eqn:ssprofile} are less than 1, i.e. $a(v+q)/(uq+pv+pq)<1$ (for $s^{A}$) and $au/(uq+pv)<1$ (for $s^{B}$). Thus, having rates which satisfy these two inequalities simultaneously is a necessary and sufficient condition for the system to be in steady phase.
\item Away from the left boundary, $s^{A}_{i}$ and $s^{B}_{i}$ approach asymptotic values that satisfy the relation $s^{A}_{i}/s^{B}_{i} = v/u$ which corresponds to zero interconversion current at site $i$.

\item The characteristic length ($l_{D}$) associated with this (exponential) approach is given by 
\begin{equation}
 l_{D} = \left[\ln \left(\frac{1}{\lambda}\right)\right]^{-1} 
\label{eqn:decay_length}
\end{equation}
 which is $(v/q + u/p)^{-1}$ to first order in interconversion rates. This first order term has the following interpretation. $1/p$ is the time over which a single hopping event of an A particle takes place at each site in the lattice. $u$ is the rate of conversion for an A particle. Thus, $u/p$ is the probability that a conversion event (of an A particle) takes place in this time. The term $v/q$ has a similar interpretation. Thus, $(v/q + u/p)$ is an estimate of the average number of interconversions taking place in the system in a time interval over which or before hopping occurs. As reflected in the expression for $l_{D}$ (and discussed in section \ref{sec:results}), when this number is large, $s^{A}$ and $s^{B}$ approach their asymptotic values over just a few sites from the left boundary. Conversely, for interconversion rates that are proportional to $1/L$, the length $l_{D}$ is of order $L$ and the gradients in $s^{A},s^{B}$ extend over the bulk of the lattice. ($\ensuremath{\mathcal{O}\!\left(1/L\right)}$ interconversion rates are also studied in \cite{frey,juhasz}.)
\end{enumerate}

\subsection{Growing Phases:}
\begin{figure}
\centering
\includegraphics[width=0.48\textwidth]{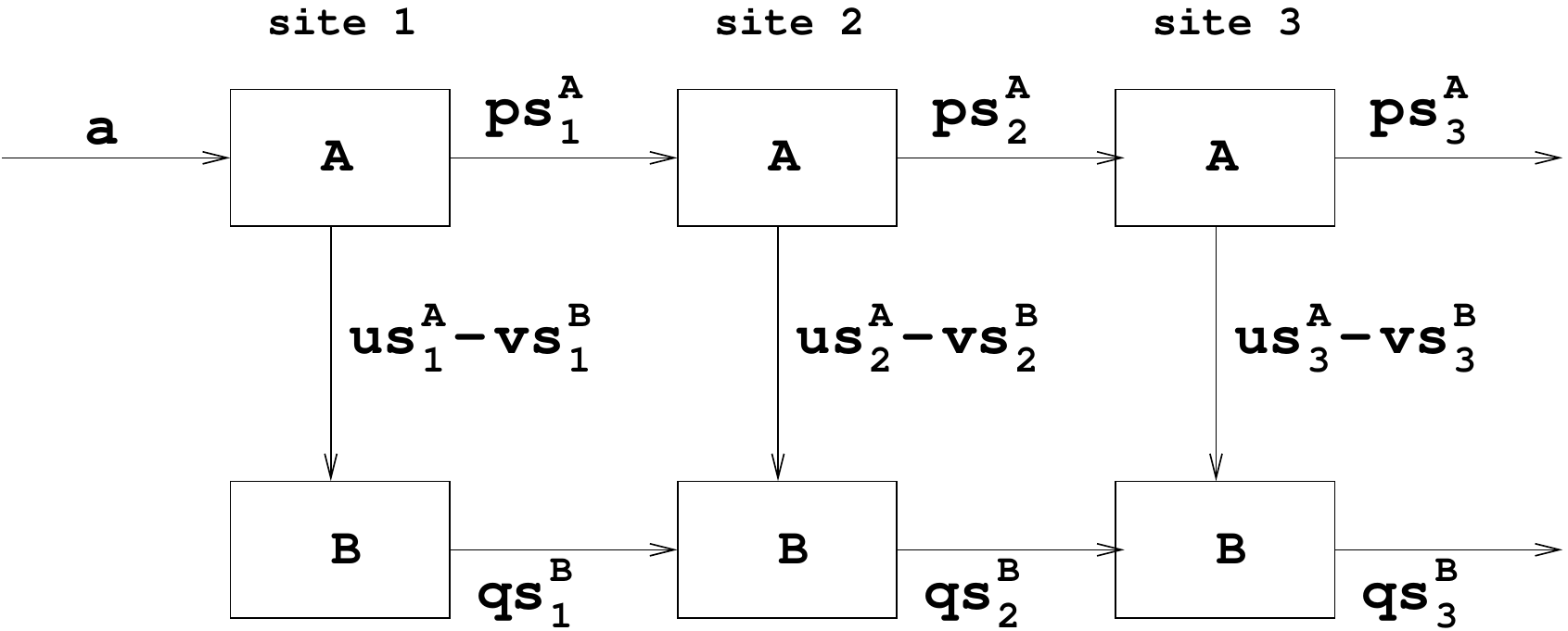}
\caption{Each site can be thought of as having A and B compartments. Exchange of particles takes place between compartments at a give site, i.e. through interconversion currents (represented by vertical arrows). Exchange of particles takes place between sites through hopping currents (represented by horizontal arrows).}
\label{network}
\end{figure}

\paragraph*{} To study the structure of the growing phases in the system, we use Fig. \ref{network} which is a diagrammatic version of Eqs. \eqref{eqn:currenteqsitei} and \eqref{eqn:currenteqsite1A}. Each site can be thought of as having A and B compartments. The figure shows particle currents between the various compartments, with vertical arrows representing intra-site interconversion currents and horizontal arrows the inter-site hopping currents. The currents into and out of any compartment can either balance each other (resulting in a steady state for the compartment) or the in current can be more than the out current (resulting in a pile-up). Evidently, the out current cannot exceed the in current at any site as this would eventually lead to negative mass. This simple observation can be used to establish the following result: \\ \emph{In both steady and growing phases, if $q\neq 0$, then $s_{i+1}^{A} < s_{i}^{A}$ and $s_{i+1}^{B} \geq s_{i}^{B}$  $\forall i$.} Consequently, $us_{i+1}^{A}-vs_{i+1}^{B} < us_{i}^{A}-vs_{i}^{B}$, i.e. interconversion current keeps decreasing as we move right, asymptotically approaching zero.
\paragraph*{}To prove this result, we note that if $q \neq 0$, then at site 1, the interconversion current, being the only `in' current for compartment B, is always positive, i.e. from A to B. This implies that $s_{2}^{A} < s_{1}^{A}$ and $s_{2}^{B} \geq s_{1}^{B}$. We can prove this by ruling out the other three possibilities which are: 
\\(i) $s_{2}^{A} \geq s_{1}^{A}$, $s_{2}^{B} \geq s_{1}^{B}$: This  would imply that either the net current out of site 2 is more than the net current into it which is not possible; or if $s_{1}^{A}=s_{2}^{A}$ and $s_{1}^{B}=s_{2}^{B}$, then $us_{2}^{A}-vs_{2}^{B}=us_{1}^{A}-vs_{1}^{B}>0$, i.e. there is a net interconversion current out of A. Since the in and out hopping currents for A are equal ($s_{1}^{A}=s_{2}^{A}$), this means there is a net out current from compartment A which is not possible.  
\\(ii) $s_{2}^{A} < s_{1}^{A}$, $s_{2}^{B} < s_{1}^{B}$: Since $s_{1}^{A}$ and $s_{1}^{B}$ can at most be equal to $1$, this implies that $s_{2}^{A},s_{2}^{B}<1$, i.e. site 2 is in steady state and currents for both compartments balance. To satisfy current balance for A with $s_{2}^{A} < s_{1}^{A}$, interconversion current must be from A to B (see Fig. \ref{network}) but for current balance for B with $s_{2}^{B} < s_{1}^{B}$, it must be in the opposite direction. This leads to a contradiction. 
\\(iii) $s_{2}^{A} \geq s_{1}^{A}$, $s_{2}^{B} < s_{1}^{B}$: This implies that $us_{2}^{A}-vs_{2}^{B} > us_{1}^{A}-vs_{1}^{B}>0$. However (from Fig. \ref{network}), to have $s_{2}^{A} \geq s_{1}^{A}$, the interconversion current has to be from B to A or at least has to be zero. This leads to an inconsistency. 
\\Thus, the only possibility is that  $s_{2}^{A} < s_{1}^{A}$ and $s_{2}^{B} \geq s_{1}^{B}$. In the same way, the proof can be extended to all subsequent sites. It may be noted that we have not assumed the existence of steady state {\bf i.e. $d\langle m^{A,B}\rangle /dt=0$,} anywhere in this argument. Thus, the inequalities hold for the growing phase also. The $q = 0$ case has many sub cases and can be analysed in a similar way. In the rest of the discussion, it is assumed that $q\neq 0$.

\paragraph*{}
The above inequalities for the occupation probabilities have the following implications:
 \begin{enumerate}
  \item Even if a pile-up of A exists on site 1 ($s_{1}^{A}=1$), no subsequent site can have a pile-up of A (for $q \neq 0$) as $s_{i}^{A} < s_{1}^{A}$  $\forall i>1$. Thus, an A pile-up would be found only at site 1, if at all.
\item The occupation probability $s^{B}$ increases from left to right. If  $s^{B}$ touches 1 at some site $k$, then $s_{i}^{B}=1$    $\forall i>k$ as $s_{i}^{B} \geq s_{k}^{B}$    $\forall i>k$. Thus, there are \emph{B pile-ups at all sites right of $k$}. 
 \end{enumerate}

\paragraph*{}
It also follows from Eq. \eqref{eqn:currenteqsitei} that  when B pile-ups exist, then at long times, the mass of B grows linearly with time at all sites with pile-ups (except at site $k$, the earliest site with a B pile-up, where it may grow as $\sqrt{t}$). 
Moreover, if $m_{k+1}^{B} \sim ct$, then $m_{k+2}^{B} \sim \left(p/(p+u)\right) ct$, $m_{k+3}^{B} \sim \left(p/(p+u)\right)^{2} ct$ etc. Thus, after sufficiently long times, a snapshot  of the lattice at any instant would show an exponential decay in space in the mass profile of B, site $k+1$ onwards, with a decay length given by $\left[\ln \left((p+u)/p\right)\right]^{-1}$. The location of the site $k$ depends on the parameters and can even be tuned to be at site 1. Left of site $k$, the system exists in steady state (except for a possible A pile-up on site 1) while right of it, the system exists in a growing phase.

\begin{figure}
\includegraphics[width=0.48\textwidth]{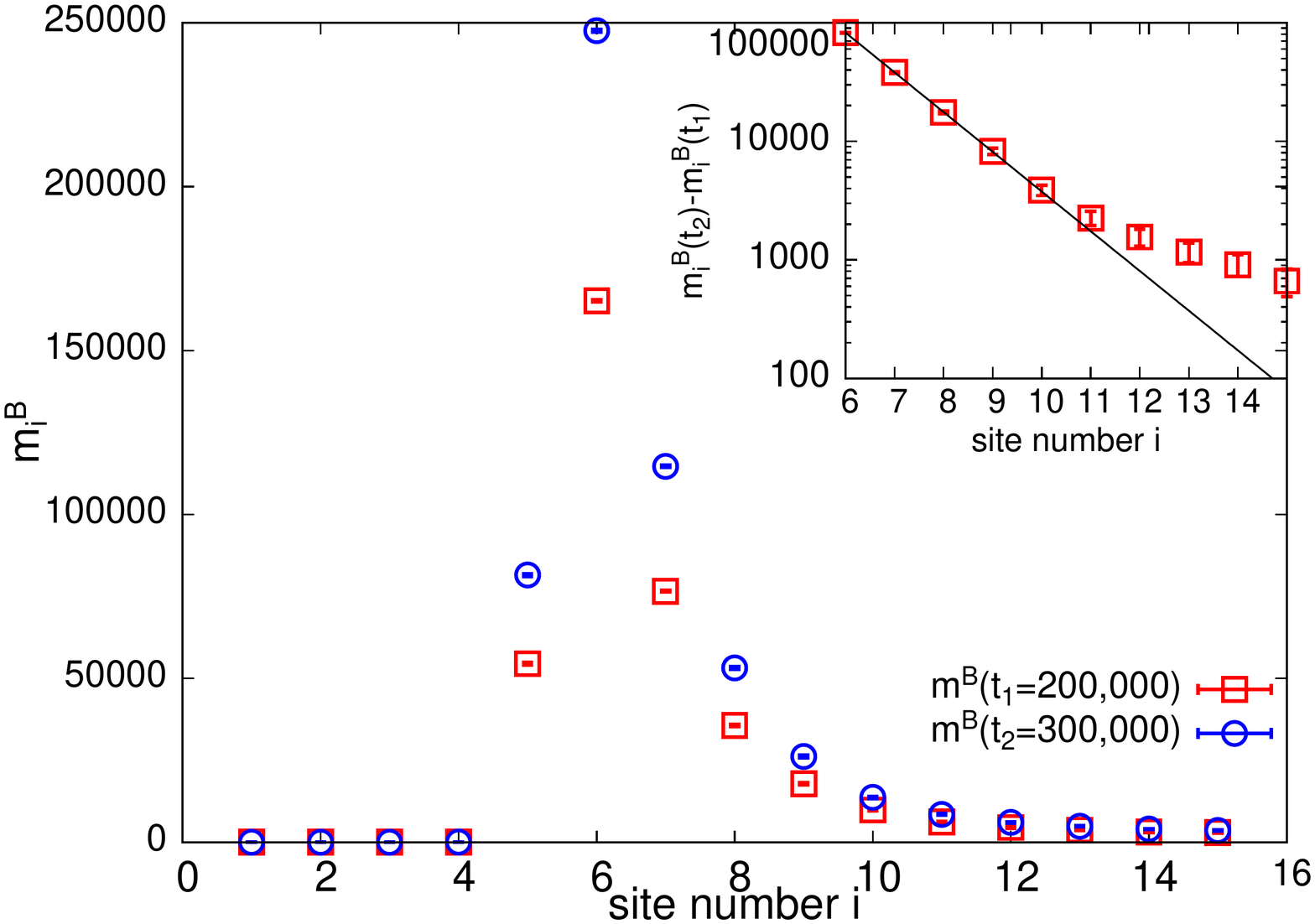}
\caption{Results of Monte Carlo simulations for $a=1.3$, $u=1$, $v=0.33$, $p=0.86$, $q=1$. Snapshots of the lattice showing $m^{B}$ profile at $t_{1}=200000$ (squares) and $t_{2}=300000$ (circles) ($t$ in units of 500 MC Steps). Inset: Semi-log plot of $m^{B}_{i}(t_{2})-m^{B}_{i}(t_{1})$ vs $i$. Solid line shows the analytical prediction.}
\label{profile}
\end{figure}

\paragraph*{}
The system was also studied numerically by doing Monte Carlo simulations for small lattices, typically $L=20$. Since later sites cannot affect earlier sites, this is the same as doing a simulation with large $L$ but keeping track of only the first few sites. Numerical results are presented for a choice of parameters for which the theory predicts the existence of steady state at the first four sites, and pile-ups of B site 5 onwards. Figure \ref{profile} shows  snapshots of the B profile of the lattice at two times $t_{1}=200,000$ and $t_{2}=300,000$ (where t is in units of 500 Monte Carlo (MC) steps). In agreement with the analytical prediction, upto site 4, $m^{B}$ reaches a finite time-independent value, while growth of $m^{B}$ starts occurring from site 5, is fastest at site 6 and becomes slower as we go further from the boundary. The inset of Fig. \ref{profile} shows  $m^{B}(t_{2})-m^{B}(t_{1})$ as a function of site number in a semi-log plot. The solid line is the analytically expected exponential decay. There is good agreement between the solid line and the data points until site 10 but not after that. For sites after site $10$, the long time behaviour, i.e. linear growth does not set in  during the run time of the simulation ($1.5\times 10^{8}$ MC steps). Such large relaxation times were the primary reason for choosing small system sizes in simulations.

\paragraph*{}
Thus, to sum up, for only rightward hopping ($\gamma=1$), with injection of A particles at site $1$, $s^{A}$ decreases monotonically while $s^{B}$ increases monotonically with site number. The system may attain stationarity or it may exist in a growing phase characterised by `pile-ups'. An A pile-up can occur only at the first (boundary) site (for $q\neq0$). B pile-ups occur in the bulk; in fact, it is possible to have steady state existing near the right boundary and pile-ups of B further on (as in Fig \ref{profile}). By making interconversion much slower than hopping ($u/w + v/q \sim \ensuremath{\mathcal{O}\!\left(1/L\right)}$), both the steady state and the pile-up region can be made macroscopically large ($\ensuremath{\mathcal{O}\!\left(L\right)}$). If a site has a pile-up of B, then all sites right of it also have B pile-ups. These pile-ups grow as $\sim t$. In the pile-up region, at any instant, the spatial profile of $m^{B}$ shows an exponential decay.

\paragraph*{}
When $q=0$, i.e. B particles cannot move, there are two important differences. Steady state on site 1 implies a steady state behaviour for the whole system, and it is not possible to have a region of steady state followed by a region of pile-ups. Secondly, $q=0$ is the only surface in parameter space on which \emph{pile-ups of A need not be localised at site 1}. If there exists a pile-up of A on site 1, then as long as $u\leq v$, there are pile-ups of A on all subsequent sites. These pile-ups grow as $\sqrt{t}$ with time. 

\section{Continuum limit}
\label{sec:continuum}
\paragraph*{}
For fully asymmetric hopping ($\gamma=1$), steady phase occupation probabilities were obtained by solving Eqs. \eqref{eqn:currenteqsitei}, \eqref{eqn:currenteqsite1A}. For any other $\gamma$ however, a given site is affected by both its left and right neighbours and the resulting time evolution equations for $<m_{i}^{A}>$ and $<m_{i}^{B}>$, (similar to Eqs. \eqref{eqn:currenteqsitei}, \eqref{eqn:currenteqsite1A}) yield complicated recursion relations for the occupation probabilities which are difficult to solve. Thus, to analyse the system for any general asymmetry, we assume the total number of sites $L$ in the system to be large, and take a continuum limit for the lattice, thereby going from the integer valued site number $i$ to the real valued position co-ordinate $x$, defined as $x=i/L$. By this definition, $x$ can take on real values between $0$ and $1$. In the continuum limit, 
\begin{equation}
 s_{i\pm 1}^{A,B} = s^{A,B}(x) \pm \frac{1}{L} \frac{\partial s^{A,B}}{\partial x} +\frac{1}{2L^{2}} \frac{\partial^{2} s^{A,B}}{\partial x^{2}} + \hspace{2pt} ...
\label{eqn:continuum}
\end{equation}
 
\subsection{$\gamma=1$}
\label{sec:gamma1}
\paragraph*{}
First we would like to check whether the steady phase profile given by Eq. \eqref{eqn:ssprofile} is recovered in the continuum limit. Substituting from Eq. \eqref{eqn:continuum} into Eqs. \eqref{eqn:currenteqsitei} and \eqref{eqn:currenteqsite1A}, setting time derivatives equal to zero and retaining only leading order terms in $1/L$, we get:
\begin{subequations}
\label{eqn:asym_cont}
 \begin{equation}
-p\frac{ds^{A}}{dx}+\tilde{v}s^{B}-\tilde{u}s^{A}=0
 \end{equation}
\begin{equation}
-q\frac{ds^{B}}{dx}+\tilde{u}s^{A}-\tilde{v}s^{B}=0
\end{equation}
\end{subequations}
 along with the boundary conditions $s^{A}(x=0) = a/p$ and $s^{B}(x=0) = 0$. Here, $\tilde{u}$ and $\tilde{v}$ refer to the rescaled rates $\tilde{u} = uL$ and $\tilde{v} = vL$. In Eq. \eqref{eqn:asym_cont} the full derivative with respect to $x$ has been used instead of the partial derivative as the steady state occupation probabilities are functions of $x$ only. The solution of Eq. \eqref{eqn:asym_cont} is given by:
 \begin{subequations}
 \label{eqn:asym_cont_profile}
 \begin{equation}
  s^{A}(x)=\frac{a\tilde{v}}{\tilde{u}q+\tilde{v}p}\left[1+ \frac{\tilde{u}q}{\tilde{v}p} \exp\left(-\frac{x}{\tilde{l}}\right)\right] 
\end{equation}
 \begin{equation}
  s^{B}(x)=\frac{a\tilde{u}}{\tilde{u}q+\tilde{v}p}\left[1 -\exp\left(-\frac{x}{\tilde{l}}\right)\right]
 \end{equation}
    \end{subequations}
  with $\tilde{l} = \left(\tilde{v}/q + \tilde{u}/p\right)^{-1} = (1/L) \left(v/q + u/p\right)^{-1}$. A comparison with the decay length computed for the discrete lattice (in Eq. \eqref{eqn:decay_length}) shows that apart from the length rescaling factor $1/L$,  $\tilde{l}$ is just the first order term in the expansion of the exact decay length $l_{D}$ in terms of $(v/q + u/p)$. This first order term arises because in retaining only the leading order term in Eq. \eqref{eqn:continuum}, we are making the assumption that $s^{A,B}(x)$ are slowly varying functions of $x$ which in turn requires that interconversions are much slower than hopping, i.e. $(v/q + u/p) << 1$.

\subsection{$ \gamma =1/2 $}
\label{sec: gamma1/2}
\paragraph*{}
The other limiting case is $\gamma=1/2$ or perfectly symmetric hopping. As before, current equations similar to Eqs. \eqref{eqn:currenteqsitei} and \eqref{eqn:currenteqsite1A} can be written for the discrete lattice, which on taking the continuum limit, yield the following two coupled differential equations:

\vspace{20pt}
\begin{subequations}
\label{eqn:sym_cont}
 \begin{equation}
\frac{\partial <m^{A}(x,t)>}{\partial t} = \frac{p}{2}\frac{\partial^{2}s^{A}}{\partial x^{2}}+\tilde{v}s^{B}-\tilde{u}s^{A}
 \label{eqn:sym_cont_A}
\end{equation}
\begin{equation}
\frac{\partial <m^{B}(x,t)>}{\partial t} = \frac{q}{2}\frac{\partial^{2}s^{B}}{\partial x^{2}}+\tilde{u}s^{A}-\tilde{v}s^{B}
\label{eqn:sym_cont_B}
\end{equation}
\end{subequations}
Here the rescaled interconversion rates are  $\tilde{u}=uL^{2}$ and $\tilde{v}=vL^{2}$. The boundary conditions are given by $s^{A}(0) = 2a/p$, $s^{B}(0) = 0$ and $s^{A}(1) = s^{B}(1) = 0$. The boundary condition at $x=1$ just reflects the fact that the right boundary acts as a sink, as particles cannot return to the system once they leave from site $L$. Equation \eqref{eqn:sym_cont} can be solved for steady phase by setting time derivatives equal to zero, taking second derivative of Eq. \eqref{eqn:sym_cont_A} with respect to $x$ and eliminating the $s^{B}$ terms by using Eq. \eqref{eqn:sym_cont_B} and the original Eq. \eqref{eqn:sym_cont_A}. This gives an equation which is the second derivative with respect to $x$ of a second order differential equation in $s^{A}$. The solution of this equation gives the following steady phase spatial profiles of $s^{A}$ and $s^{B}$:

\begin{subequations}
\label{eqn: sym_profile}
\begin{equation}
  s^{A}(x)=\frac{2a\tilde{v}}{\tilde{u}q+\tilde{v}p}\left[(1-x) + \frac{\tilde{u}q}{\tilde{v}p} \left(\frac{\sinh\left(\sqrt{\tilde{\eta}}(1-x)\right)}{{\sinh\left(\sqrt{\tilde{\eta}}\right)}}\right)\right]
\end{equation}
\begin{equation}
 s^{B}(x)=\frac{2a\tilde{u}}{\tilde{u}q+\tilde{v}p}\left[(1-x) - \left(\frac{\sinh\left(\sqrt{\tilde{\eta}}(1-x)\right)}{{\sinh\left(\sqrt{\tilde{\eta}}\right)}}\right)\right]
\end{equation}

\end{subequations}
 where $\tilde{\eta} = 2\left(\tilde{v}/q+\tilde{u}/p\right)$

\begin{figure}
\subfigure [] {
\includegraphics[width=0.48\textwidth]{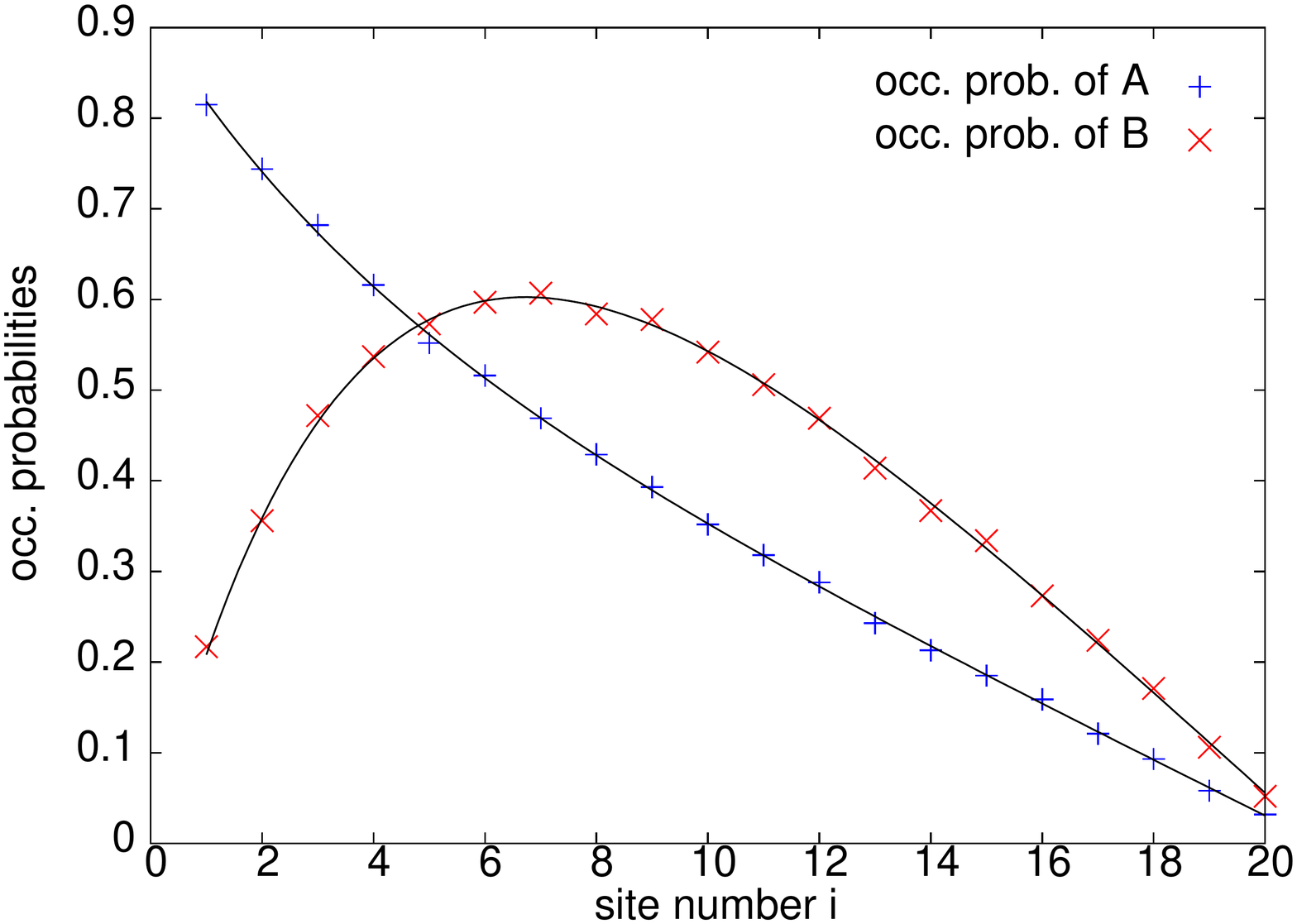}
\label{profile_sym}}
\subfigure[]{
\includegraphics[width=0.48\textwidth]{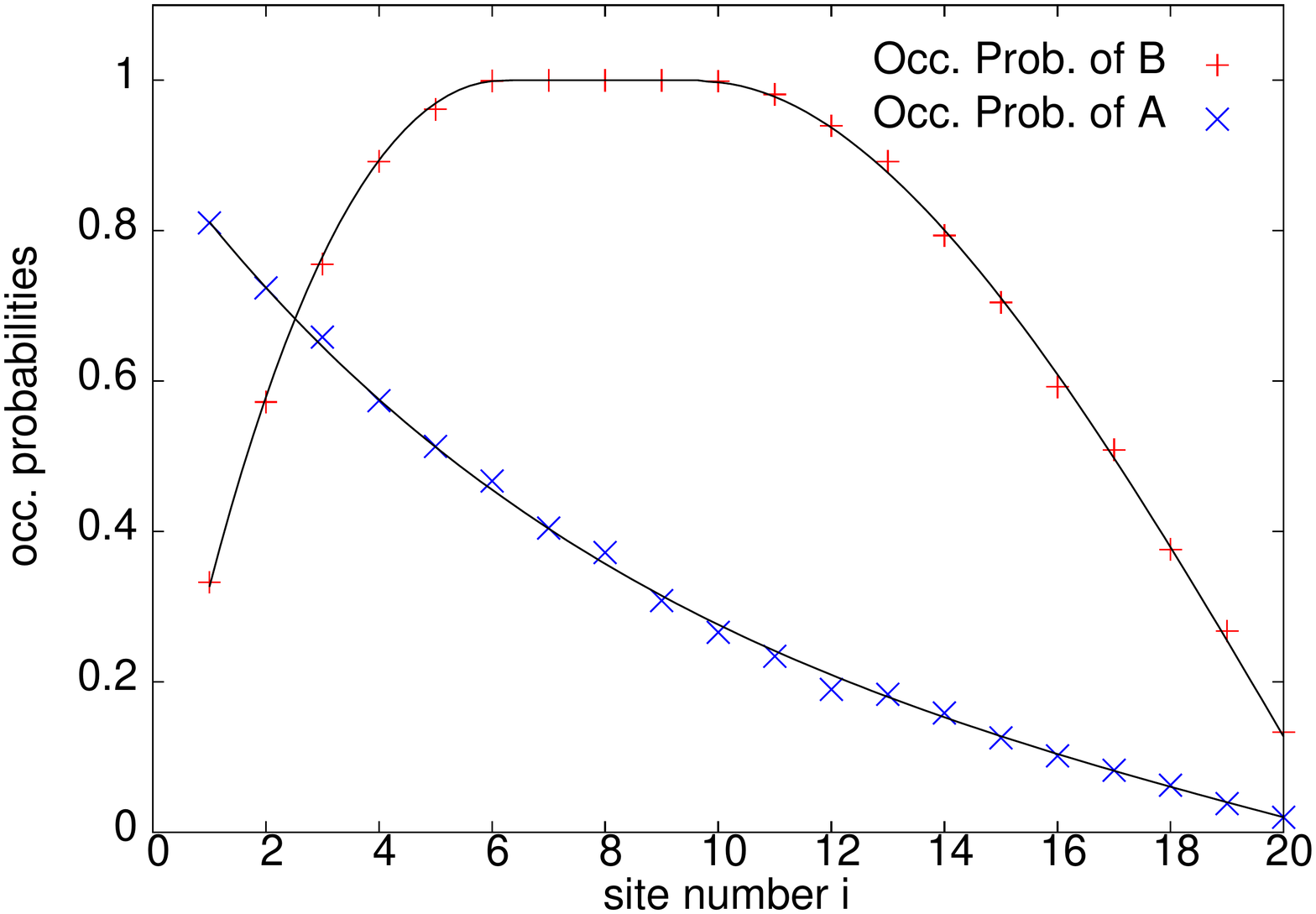}
\label{pileups_profile}}
\caption{Occupation probabilities $s_{i}^{A}$ and $s_{i}^{B}$ as a function of site number $i$  \newline (a)  in a steady phase (with $a=1.3$, $u=0.02$, $v=0.01$, $p=2.2$, $q=0.5$). \newline (b) in a growing phase (with $a=1.0$, $u=0.0136$, $v=0$, $p=2.2$, $q=0.3$). \newline Data points obtained from Monte Carlo simulations seem to agree well with analytical predictions (solid lines).}
\end{figure}
 
\paragraph*{}
Figure \ref{profile_sym} shows the spatial profile of $s^{A}$ and $s^{B}$ in steady phase for a certain choice of parameters. Clearly, steady phase profiles (Eq. \eqref{eqn: sym_profile}) obtained from the continuum approximation tally well with the Monte Carlo results.  

\paragraph*{}
The spatial profiles of $s^{A}$ and $s^{B}$ in Fig. \ref{profile_sym} give an indication of the nature of the growing phases of the system. As in the asymmetric case with $q \neq 0$, since $s^{A}$ is always a decreasing function of $x$, it follows that an A pile-up can exist only on the first site. If an A pile-up exists at site 1, then we use the boundary condition $s^{A}(x=1/L) = 1$ instead of the boundary condition for $s^{A}(x=0)$ in Eq. \eqref{eqn:sym_cont} to solve for the steady state of the rest of the system. From Fig. \ref{profile_sym} we can also deduce that if the maximum in the $s^{B}(x)$ spatial profile crosses 1, the system cannot exist in a steady phase but must show pile-ups of B. Moreover, these B pile-ups occur in some central region of the lattice flanked by regions of steady state on either side. To test this prediction, simulations were done with parameters for which $s^{B}(x)$, as obtained from Eq. \eqref{eqn: sym_profile}, crosses 1 in some range of $x$ values. Figure \ref{pileups_profile} shows the occupation probability profiles for such a choice of parameters. As expected, the region of pile-ups (with $s^{B} = 1$) exists in the middle of two steady state regions. Appendix \ref{appendixB} presents the details of how we analytically calculate the spatial boundaries of the pile-up region as well as the occupation probability profiles which are shown as solid lines in Fig. \ref{pileups_profile}. The main assumption involved in this calculation is that $\partial s^{B}/\partial x$ vanishes at the boundaries of the pile-up region.

\subsection{$ 1/2 < \gamma < 1 $}
\paragraph*{}
Away from the limits of pure drift ($\gamma = 1$) or pure diffusion ($\gamma = 1/2$), current equations similar to Eq. \eqref{eqn:sym_cont} can be written with partially asymmetric hopping rates ($1/2 < \gamma < 1$). These equations also have terms with first derivatives of $s^{A}$ and $s^{B}$ with respect to $x$, and can be solved in the same manner as Eq. \eqref{eqn:sym_cont}, giving the following expressions for the steady phase occupation probabilities $s^{A}(x)$ and $s^{B}(x)$:

\begin{subequations}
\label{eqn: p_asym_profile}
\begin{equation}
\begin{split}
s^{A}(x) & =\frac{a\tilde{v}}{\gamma(\tilde{u}q+\tilde{v}p)}\Vast[\frac{1-\exp(-2\tilde{\xi}(1-x)) }{1-\exp(-2\tilde{\xi})}\\
& + \frac{\tilde{u}q}{\tilde{v}p}\exp(\tilde{\xi}) \left(\frac{\sinh\left(\sqrt{\tilde{\eta} + \tilde{\xi}^{2}}(1-x)\right)}{\sinh\left(\sqrt{\tilde{\eta} + \tilde{\xi}^{2}}\right)}\right) \Vast]
\end{split}
\end{equation}
\begin{equation}
\begin{split}
 s^{B}(x) & =\frac{a\tilde{u}}{\gamma(\tilde{u}q+\tilde{v}p)}\Vast[\frac{1-\exp(-2\tilde{\xi}(1-x)) }{1-\exp(-2\tilde{\xi})}\\
& -\exp(\tilde{\xi}) \left(\frac{\sinh\left(\sqrt{\tilde{\eta} + \tilde{\xi}^{2}}(1-x)\right)}{\sinh\left(\sqrt{\tilde{\eta} + \tilde{\xi}^{2}}\right)}\right)\Vast]
\end{split}
\end{equation}
\end{subequations}
 where $\tilde{u} = uL^{2}$, $\tilde{v} = vL^{2}$, $\tilde{\eta} = 2\left(\tilde{v}/q+\tilde{u}/p\right)$ and $\tilde{\xi} = (2\gamma - 1)L$

\paragraph*{}
Equation \eqref{eqn: p_asym_profile} is also valid for $0 < \gamma <1/2$. However, if particles have a net drift back towards the source, only a small region close to $x=0$ is occupied. Thus, we confine our analysis to the more interesting case $1/2 < \gamma <1$. Strictly at the two limits, $\gamma = 1/2$ and $\gamma = 1$, Eq. \eqref{eqn: p_asym_profile} is not valid. However, taking the limit $\gamma \rightarrow 1/2$, or equivalently $\tilde{\xi}\rightarrow 0$ for Eq. \eqref{eqn: p_asym_profile} gives back the profiles of Eq. \eqref{eqn: sym_profile}. At $\gamma = 1$, a second order description cannot be used as there is only one boundary condition for $s^{A}$ (or $s^{B}$).

\begin{figure}
\vspace{-5mm}
\centering 
\includegraphics[width=0.48\textwidth]{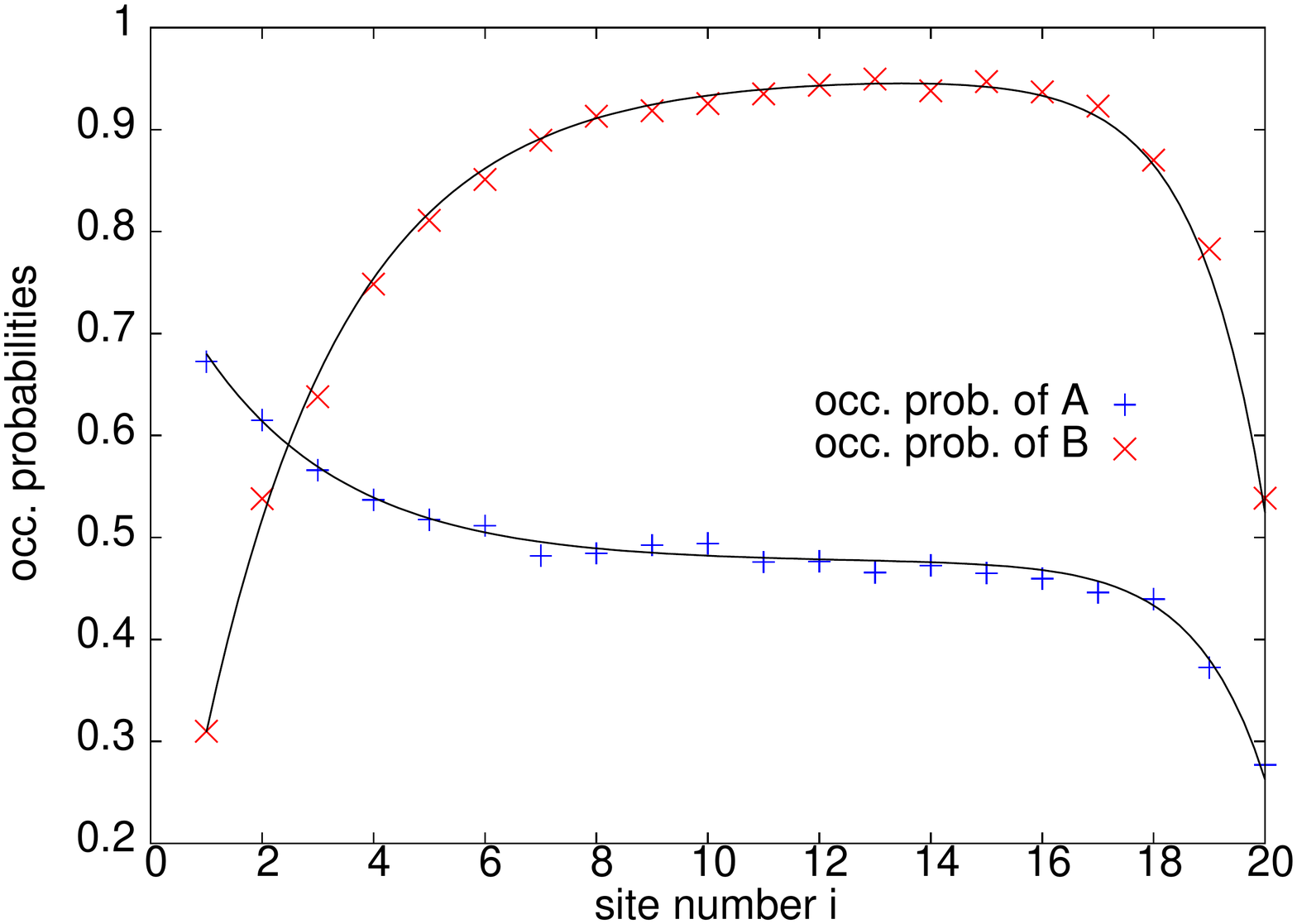}
\vspace{-8mm}
\caption{$s^{A}_{i}$ and $s^{B}_{i}$  for $a=1.2$, $u=0.2$, $v=0.1$, $p=2.2$, $q=0.7$ and $\gamma = 0.7$. For these parameters, $\tilde{\eta}=103$ and $\tilde{\xi}=8.4$. Solid lines are plots of Eq. \eqref{eqn: p_asym_profile} with $x$ replaced by $i/L$.}

\label{partial_asym_ss}
\end{figure}

\paragraph*{}
The spatial variation of $s^{A}(x)$ and $s^{B}(x)$ is governed by the rescaled variables $\tilde{\eta}$ and $\tilde{\xi}$. If $\tilde{\eta} \sim \ensuremath{\mathcal{O}\!\left(L^{2}\right)}$ and $\tilde{\xi} \sim \ensuremath{\mathcal{O}\!\left(L\right)}$, this variation is confined to the boundary regions, eventually becoming a discontinuity in the limit $L\rightarrow\infty$. If $\tilde{\eta} \sim \ensuremath{\mathcal{O}\!\left(1\right)}$ and $\tilde{\xi} \sim \ensuremath{\mathcal{O}\!\left(1\right)}$, the gradients due to the two boundaries extend into the whole lattice and together, determine the behaviour of the system in a complex way. For large but not infinite values of $\tilde{\eta}$ and $\tilde{\xi}$ ($\tilde{\eta} \sim \ensuremath{\mathcal{O}\!\left(100\right)}$, $\tilde{\xi} \sim \ensuremath{\mathcal{O}\!\left(10\right)}$), gradients near the two boundaries can be treated independently and yet are not mere boundary discontinuities. This situation is shown in Fig. \ref{partial_asym_ss}. For such values of $\tilde{\eta}$ and $\tilde{\xi}$ and sufficiently large $L$, two distinct length scales emerge, one characterizing the variation near $x = 0$ and the other near $x = 1$. These length scales are, respectively:
\begin{equation}
 \tilde{l}_{1} = \left[\sqrt{\tilde{\eta} + \tilde{\xi}^{2}} - \tilde{\xi}\right]^{-1}, \qquad \tilde{l}_{2} = \left[2\tilde{\xi}\right]^{-1}
\end{equation}

\paragraph*{}
The length $\tilde{l}_{1}$ is the distance from $x=0$ beyond which A and B are effectively decoupled (no net interconversion). $\tilde{l}_{1}$ is small for fast interconversions (high values of $\tilde{\eta}$), which has been rationalized earlier in section \ref{sec:results}. $\tilde{l}_{1}$ also decreases as $\tilde{\xi}$ decreases, i.e. with decrease in hopping asymmetry.
\paragraph*{}
The effect of the absorbing right boundary (at $x=1$) extends over a length which is given by $\tilde{l}_{2}$. This can be seen in the following way. Since A and B are effectively decoupled beyond $\tilde{l}_{1}$, the spatial variation of $s^{A}(x)$ (and similarly $s^{B}(x)$) beyond this length is described by a drift diffusion equation ($p/2(\partial^{2}s^{A}/\partial x^{2}) - p\tilde{\xi}(\partial s^{A}/\partial x) \simeq 0$) along with the boundary condition $s^{A}(x=1)=0$. The diffusion length associated with this equation is $[2\tilde{\xi}]^{-1}$ which is just $\tilde{l}_{2}$. 
\paragraph*{}
As $\tilde{l}_{1}$ and $\tilde{l}_{2}$ increase, the variations near the two boundaries can no longer be treated independently and the middle region of nearly constant $s^{A}(x)$ and $s^{B}(x)$ seen in Fig. \ref{partial_asym_ss} disappears. Perfectly symmetric hopping, for which $\tilde{l_{2}}$ diverges, is an extreme case of this.
\section{Multi-Species Model}
\label{sec:3species}
\begin{figure}
\vspace{-5mm}
\centering 
\includegraphics[width=0.48\textwidth]{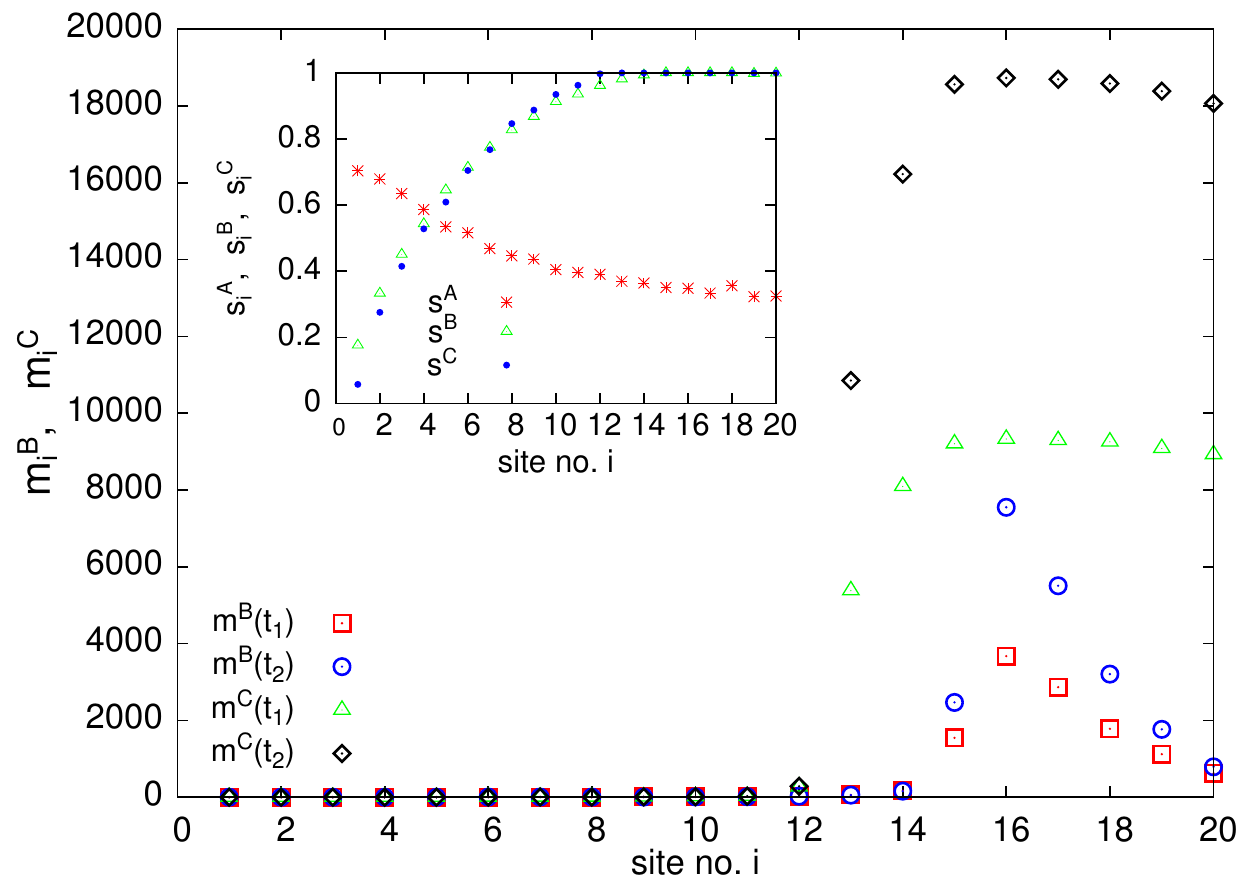}
\caption{Results of Monte Carlo simulations for a phase with pile-ups of both B and C ($\gamma=1$). Parameters: $a=5$, $w=6$, $q=2$, $r=0.3$, $u=0.7$, $v=0.2$, $k=0.65$, $l=0.62$. Data points show $m^{B}_{i}$ and $m^{C}_{i}$  at two time intervals $t_{1}=5 \times 10^{6}$ and $t_{2}=10^{7}$. Pile-ups of C occur earlier in the lattice (site 13 onwards) than pile-ups of B (site 15 onwards). Inset: $s_{i}^{A}$, $s_{i}^{B}$, $s_{i}^{C}$ at long times.}
\label{3sp}
\end{figure}

\paragraph*{}
The two-species model can be extended quite easily to more species. For example, consider the three-species generalization. Particles of type A are injected at the left boundary. An A particle can convert to B and vice versa; in addition, B particles can convert to C (at rate $k$) and C to B (at rate $l$). Note that direct interconversion between A and C is not allowed. 
Also, as before, a single particle of any species can chip off a site (with rates $p,q,r$ for species A,B,C respectively) and move, in general, in a driven diffusive way.
\paragraph*{}
The model can be analysed in exactly the same way as the two-species model by writing equations involving particle currents. The analysis of the three-species model yields the following results:
\begin{enumerate}
\item While $s^{A}$ decreases monotonically as a function of $x$, $s^{B}$ and $s^{C}$ both show qualitatively the same behaviour of the spatial profile, i.e. both increase monotonically with $x$ in the driven case and have a peak in the diffusive case. This indicates that in the three-species model, depending on the rates, pile-ups of B or C or of both can occur in the bulk of the lattice and a pile-up of A only on the first site.
\item Interestingly, although interconversion takes place in the order $A\rightleftharpoons B\rightleftharpoons C$, pile-ups need not appear in the same order in space. For example, C pile-ups may be found to the left of B pile-ups (i.e. closer to the left boundary) as shown in Fig \ref{3sp}.
\end{enumerate}

\paragraph*{}
Thus, in the multi-species model with sequential interconversion, in different regions of parameter space, the system may have pile-ups of one, a few or all (except A) species of particle. Pile-ups of different species may occur in different regions of the system. 
\section{Conclusion}
\label{sec:conclusion}
\paragraph*{}
In this paper, motivated by the phenomenology of traffic in the Golgi apparatus of the cell, we have studied a stochastic two-species model with boundary injection of type A, interconversion between types A and B, and transport of both species through the lattice (in general, in a driven diffusive way) by chipping of one particle at a time. We found that depending on the rates of various processes, the system may either eventually attain steady state or show unbounded growth of mass at all times. Pile-ups (as defined in the text) may be composed of A particles or B particles or both. Unlike in translationally invariant systems, the phases in our model have interesting spatial structure. Generically, in growing phases, a part of the lattice attains steady state while other regions show unbounded growth of mass (pile-ups). 
\paragraph*{}
Below, we comment on some possible extensions of the model:
\begin{enumerate}
 \item If stack movement also occurs (as in the more general model defined in section \ref{sec:model}), there are \emph{no growing phases} and the average mass in the system always attains a constant finite value. This may be argued as follows. If $j_{in}$ and $j_{out}$ are the in and out currents at a site, then the average mass $m$ at the site grows as $dm/dt=j_{in}-j_{out}$. In the `chipping only' case, $j_{out}=ps^{A}$ (assuming only one type of particle). If $j_{in}>p$, then in and out currents cannot balance, and $m$ grows indefinitely. On the other hand, with stack movement, $j_{out}$ which is given by $ps^{A}+Dm^{A}$, can always balance $j_{in}$, resulting in zero growth rate or constant mass. However, if the rate of stack movement is mass dependent i.e. $D(m) \propto m^{-\alpha}$, then $j_{out}$ which is now $ps^{A}+D(m^{A})^{1-\alpha}$ will be bounded for $\alpha \geq 1$. In this case, pile-ups can occur and the behaviour of the system would be expected to be similar to that of the `chipping only' model. 
\item One extension of this model to higher dimensions is trivial. If injection takes place at a surface perpendicular to one of the spatial directions, and we assume periodic boundary conditions in other directions, the model is still effectively one dimensional as there are no $net$ currents between sites in the transverse direction. Generalizations to higher dimensions with other boundary conditions can be more complex. 
\item This model can also be extended quite easily to include injection of B particles at the left boundary, at a rate $b$. Now, depending on the value of the ratio $auq/bpv$, the system either has net interconversion from A to B (and behaviour similar to what has been discussed in the paper) or net interconversion from B to A (resulting in A pile-ups in the bulk etc.) or no net interconversion, i.e. no effective coupling between the two species. At the right boundary, introducing exit rates that are different from the chipping rates can also change the behaviour of the system. However, this change is only a boundary effect unless the exit rate is $\ensuremath{\mathcal{O}\!\left(1/L\right)}$ times the chipping rate. If the exit rate is small in the sense defined above, then $s^{A}$ and $s^{B}$ vary with $x$ differently from Eqs. \eqref{eqn: sym_profile}. However, growing phases with B pile-ups in the bulk are still found.
\end{enumerate}
\paragraph*{}
Finally, we comment on the possible relevance to the biological system. The model system we have studied is a simple one, but it shares the following qualitative features with the Golgi apparatus. First, there is a gradation of sizes of stacks from the source to the sink. In particular, small stacks are found close to the source while large, though growing, aggregates occur in regions farther away. Secondly, there is a gradation of `chemical' species across the system, with different spatial regions having aggregates that may be predominantly of type B and/or type C etc. These aggregates, however, show unbounded growth, which is not realistic in the context of the Golgi, which consists of large but bounded and discrete structures. As discussed above, unbounded growth is eliminated once we allow for movement of whole stacks, as in the more general model defined in Sec \ref{sec:model}, and discussed in this section.
\paragraph*{}
In the present model, we have assumed that mass transfer occurs either by chipping of a single particle or by movement of the whole stack. Other transport processes could involve the fragmentation and movement of chunks of intermediate sizes. The study of these different mass transport processes in systems with boundary injection is an interesting direction for future study.
\section*{Acknowledgements:}
\paragraph*{}
We thank A.B. Kolomeisky and D. Mukamel for useful discussions and M.R. Evans for very useful comments on the manuscript.

\appendix
\section{Possible behaviours on the first site:}
\label{appendixA}
As indicated in Section \ref{sec:first_site}, conditions for each of the four scenarios for the first site can be derived from Eq. \eqref{eqn:currenteqsite1}. They are as follows:
\begin{enumerate}
\item Steady state: If $\frac{a(v+q)}{qp+qu+pv} < 1$ and $\frac{au}{qp+qu+pv} < 1$, then steady state is reached with steady state probabilities:
\\
\\* $s^{A}=\frac{a(v+q)}{qp+qu+pv}$ and $s^{B}=\frac{au}{qp+qu+pv}$
\item Pile-up of B: If $\frac{au}{qp+qu+pv} \geq 1$ and $\frac{a+v}{u+p} < 1$, then $\langle m^{B}\rangle$ keeps growing while $\langle m^{A}\rangle$ reaches a constant value. 
\\
\\ At long times:  $s^{A}=\frac{a+v}{u+p}$, $s^{B}=1$ and $\frac{d\langle m^{B}\rangle}{dt} = \frac{ua-vp}{u+p} - q$
\item Pile-up of A: If $\frac{a(v+q)}{qp+qu+pv} \geq 1$ and $\frac{u}{v+q} < 1$, 
then $\langle m^{A}\rangle$ keeps growing while $\langle m^{B}\rangle$ reaches a constant value. 
\\
\\At long times:   $s^{B}=\frac{u}{v+q}$, $s^{A}=1$ and $\frac{d\langle m^{A}\rangle}{dt} = a-p-\frac{uq}{v+q}$
\item Pile-ups of both A and B: If $\frac{a+v}{u+p} \geq 1$ and $\frac{u}{v+q} \geq 1$, then both $\langle m^{A}\rangle$ and $\langle m^{B}\rangle$ keep growing. At long times:
\\
\\ $s^{B} = s^{A}=1$, $\frac{d\langle m^{A}\rangle}{dt} = a + v - u - p$ and $\frac{d\langle m^{B}\rangle}{dt} = u - v - q$
\end{enumerate}

\section{Method for calculating $s^{A}(x)$ and $s^{B}(x)$ for $\gamma = 1/2$ when pile-ups of B occur:}
\label{appendixB}
Let $x_{1}$ and $x_{2}$ divide the lattice into three regions such that in:
\\Region I ($0 \leq x < x_{1}$): Steady state ($s^{A}(x)<1$, $s^{B}(x)<1$);
\\Region II ($x_{1} \leq x \leq x_{2}$): Steady state of A, pile-ups of B ($s^{A}(x)<1$, $s^{B}(x)=1$);
\\Region III ($x_{2} < x \leq 1$): Steady state ($s^{A}(x)<1$, $s^{B}(x)<1$).

\paragraph*{}
In regions I and III, since steady state exists, $s^{A}(x)$ and $s^{B}(x)$ are obtained by setting the L.H.S. equal to $0$ in both of Eqs. \eqref{eqn:sym_cont_A} and \eqref{eqn:sym_cont_B} and solving them with the following boundary conditions for:
\\ Region I: $s^{A}(0)=2a/p$,  $s^{B}(0)=0$,  $s^{B}(x_{1})=1$,  $\partial s^{B}/\partial x\big|_{x=x_{1}}=0$
\\Region III: $s^{B}(x_{2})=1$,  $\partial s^{B}/\partial x\big|_{x=x_{2}}=0$, $s^{A}(1)= s^{B}(1)=0$

\paragraph*{}
In region II, substituting $s^{B}(x) = 1$ into Eq. \eqref{eqn:sym_cont_A}, and setting the left-hand-side equal to zero, we get $s^{A}(x)$ upto two unknowns. We can find $s^{A}(x_{1})$ and $s^{A}(x_{2})$ from the solutions in regions I and III respectively. Since $s^{A}(x)$ must be a continuous function, these provide the boundary conditions for the solution of $s^{A}(x)$ in region II. Thus, we have $s^{A}(x)$ and $s^{B}(x)$ for all three regions upto the two unknowns $x_{1}$ and $x_{2}$. To determine $x_{1}$ and $x_{2}$, we make the assumption that $\partial s^{A}/\partial x$ is continuous at $x_{1}$ and $x_{2}$, i.e. $\partial s^{A}/\partial x\big|_{x_{1}}^{I} = \partial s^{A}/\partial x\big|_{x_{1}}^{II}$ and $\partial s^{A}/\partial x\big|_{x_{2}}^{II} = \partial s^{A}/\partial x\big|_{x_{2}}^{III}$ which yields two transcendental equations in $x_{1}$ and $x_{2}$. For a given set of parameters, these can be solved numerically to get $x_{1}$ and $x_{2}$, Since we have expressions for $s^{A}(x)$ and $s^{B}(x)$ in terms of $x_{1}$ and $x_{2}$, this also gives the occupation probabilities in all three regions.
\paragraph*{}
This calculation involves two \emph{ad hoc} assumptions, namely, that $\partial s^{B}/\partial x$ vanishes and $\partial s^{A}/\partial x$ is continuous at the two boundaries of the pile-up region. These assumptions give results which match well with numerical data (see Fig. \ref{pileups_profile}), but they need to be investigated and adequately justified.

\end{document}